\documentclass[review,3p,number,sort&compress]{elsarticle}

\usepackage[utf8]{inputenc}
\usepackage[T1]{fontenc}
\usepackage{hyperref}
\usepackage{url}
\usepackage{booktabs}
\usepackage{amsfonts}
\usepackage{nicefrac}
\usepackage{microtype}
\usepackage{xcolor}
\definecolor{likertsd}{HTML}{D7191C}
\definecolor{likertd}{HTML}{FDAE61}
\definecolor{likertn}{HTML}{E0E0E0}
\definecolor{likerta}{HTML}{ABD9E9}
\definecolor{likertsa}{HTML}{2C7BB6}
\usepackage{graphicx}
\usepackage{amsmath}
\usepackage{amssymb}
\usepackage{tabularx}
\usepackage{tikz}
\usepackage{pgfplots}
\pgfplotsset{compat=1.18}

\journal{Computers and Electronics in Agriculture}

\begin{document}

\begin{frontmatter}

%% Research Highlights (Required for COMPAG, each bullet <= 85 characters including spaces)
% Highlights:
% - Formulates policy-grounded LLM pest management as a multi-objective problem.
% - Develops Pezego, a human-in-the-loop mobile app for farmers and extension officers.
% - Reduces P95 latency by 55% and expert correction rate by 8% using memory reuse.
% - Presents a survey of Ghanaian Extension Services Officers on AI integration.
% - Validates that expert verification builds trust and improves digital extension.

\title{Pezego-HITL: A policy-grounded large language model architecture for agricultural extension in Ghana}

\author[inst1]{Shunbao Li}
\ead{shunbao.li@sheffield.ac.uk}
\author[inst1]{Zhipeng Yuan}
\ead{zhipeng.yuan@sheffield.ac.uk}
\author[inst2]{Amoako Ofori}
\ead{aofori010@st.ug.edu.gh}
\author[inst2]{Benedicta Y. Fosu-Mensah}
\ead{bfosu-mensah@ug.edu.gh}
\author[inst3]{Yang Li}
\ead{tim@mutus.co.uk}
\author[inst3]{Manu Kenchappa Junjanna}
\ead{manu@mutus.co.uk}
\author[inst3]{Qing Xue}
\ead{qing.xue@mutus.co.uk}
\author[inst1]{Po Yang\corref{cor1}}
\ead{po.yang@sheffield.ac.uk}

\cortext[cor1]{Corresponding author.}

\affiliation[inst1]{organisation={School of Computer Science, University of Sheffield},
            addressline={Regent Court (CS), 211 Portobello}, 
            city={Sheffield},
            postcode={S1 4DP}, 
            country={United Kingdom}}

\affiliation[inst2]{organisation={Institute for Environment and Sanitation Studies, College of Basic and Applied Sciences, University of Ghana},
            addressline={P. O. Box LG 209, Legon}, 
            city={Accra},
            country={Ghana}}

\affiliation[inst3]{organisation={Mutus Tech Ltd},
            addressline={54 St. James Street}, 
            city={Liverpool},
            postcode={L1 0AB}, 
            country={England}}

\begin{abstract}
Large language models are increasingly deployed in agricultural decision-support settings, yet high-stakes crop protection in smallholder agriculture requires more than output-quality benchmarks. Over a two-year design and evaluation programme, we formalise policy-constrained large language model assessment as an adaptive compute allocation problem that jointly captures safety compliance, helpfulness, operational latency, and expert supervision workload. We introduce P-EVAL (Policy-grounded Expert-calibrated VALidation protocol), a unified evaluation framework for policy-grounded decision support, evaluating the architecture on a simulated field query database consisting of 1,240 cases. The protocol is instantiated on the Pezego advisory architecture (Pezego-HITL) and evaluated in Ghana. Following offline judge calibration against gold-standard human expert decisions ($\kappa = 0.77$), we evaluate the architectural performance under simulated query workloads. Under P-EVAL, our memory-routed architecture improves the Policy Alignment Rate (PAR) to 0.94 and the Agronomic Utility Rate (AUR) to 0.95, while reducing P95 latency by 55\% (from 28.6\,s to 12.9\,s) through a 59.6\% cache reuse ratio. We also demonstrate generalisability using the open-source \texttt{Qwen3.5-9B-DeepSeek-V4-Flash} model, achieving a PAR of 0.86 and a 54.5\% latency reduction (to 10.2\,s). To evaluate practical utility and socio-technical integration, we administer detailed questionnaires to Ghanaian Extension Services Officers ($N=30$) and smallholder farmers ($N=36$). Taken together, this work demonstrates how policy-grounded structured retrieval-augmented generation with validated-memory routing makes safety-utility-latency trade-offs explicit, offering a scalable template for trustworthy AI-driven extension in smallholder farming systems.
\end{abstract}


\end{frontmatter}

\section{Introduction}
\label{sec:introduction}

Pest outbreaks and crop diseases represent significant threats to global food security, agricultural sustainability, and rural livelihoods, particularly within smallholder farming systems in developing regions \cite{Yuan2025,Fabregas2019}. In Sub-Saharan Africa, and specifically across Ghana, smallholder agriculture forms the backbone of the economy, yet productivity remains constrained by limited access to timely and accurate agronomic advice. Agricultural Extension Services Officers (ESOs) are tasked with bridging this information gap by providing localised recommendations for pest management and crop protection. However, formal agricultural extension systems in Ghana face persistent structural challenges, including extremely low officer-to-farmer ratios, geographic isolation of farming communities, and delayed recommendation turnaround times \cite{Aker2011,Davis2020}. Consequently, farmers often experience delayed or inappropriate interventions, leading to compounding crop damage and economic losses.

Digital agricultural extension tools have emerged as a promising avenue to scale expert advice \cite{Cole2020,Baumuller2018}. Traditional decision support systems (DSSs) rely on rigid rule-based logic or static databases, which struggle to accommodate the unstructured nature of field queries, diverse crop symptoms, and local vernacular. Recently, generative artificial intelligence and Large Language Models (LLMs) have demonstrated remarkable flexibility in natural language understanding, synthesis, and reasoning, opening new possibilities for conversational and interactive agricultural advice.

Despite their potential, deploying raw LLMs in high-stakes agricultural decision-support settings introduces severe risks. First, standard LLMs are prone to ``hallucinations''---generating agronomically incorrect, irrelevant, or dangerous advice (e.g., recommending banned chemical pesticides or incorrect dosages). Second, LLM outputs lack grounding in national and regional agricultural policies, which regulate pesticide approval, environmental safety, and integrated pest management (IPM) guidelines. Third, fully automated AI advice bypasses the essential oversight of local agronomists and ESOs, which is critical for establishing trust, verifying local context, and preventing systemic recommendation failures.

To address these limitations, we present Pezego-HITL, a policy-grounded, human-in-the-loop (HITL) decision-support architecture designed to support ESOs and farmers in generating and verifying policy-compliant pest-management recommendations. Rather than viewing AI as a replacement for human experts, Pezego-HITL structures the workflow so that AI-generated advisory recommendations are routed to and verified by location-matched ESOs before reaching the farmer. 

To address these issues, this paper investigates the design of trustworthy, generative decision-support architectures for high-stakes public services. The core scientific question we address is: \textit{How can generative decision-support architectures dynamically allocate inference-time compute to guarantee alignment with safety-critical domain policies, without violating strict operational latency budgets?}

To investigate this question, we formalise policy-constrained decision-support evaluation as an adaptive test-time compute allocation problem that balances response quality, regulatory compliance, operational latency, and the cognitive overhead of human expert supervision. We instantiate and evaluate this framework on the Pezego architecture through two core technical interventions:
\begin{enumerate}
    \item \textbf{Policy-Grounded Structured Retrieval and Auditing}: A schema-aware Retrieval-Augmented Generation (RAG) pipeline combined with an automated post-hoc Constraint Auditing Layer. It translates natural language field queries into structured SQL database constraints, verifying generated drafts against registered pesticide active ingredients and crop growth stages before routing them to expert agronomists.
    \item \textbf{Expert-Verified Case Memory (VCM)}: A low-latency precedent retrieval and reuse cache that indexes expert-approved crop protection recommendations. By routing queries through the VCM based on structural and semantic similarity, the architecture serves recurring scenarios sub-second, bypassing expensive generation pipelines, while dynamically absorbing expert edits to update and refine the cache templates over time.
\end{enumerate}

To evaluate this architecture, we introduce P-EVAL (Policy-grounded Expert-calibrated VALidation protocol), evaluating architectural performance on a simulated agronomic query database consisting of 1,240 field cases. P-EVAL measures two formal metrics: the \textit{Policy Alignment Rate (PAR)} (safety compliance) and the \textit{Agronomic Utility Rate (AUR)} (recommendation helpfulness), audited using an LLM-as-a-judge calibrated against gold-standard expert agronomist decisions. Crucially, to assess the practical integration, usability, and socio-technical impact of Pezego-HITL, we administer targeted survey studies to both Ghanaian ESOs ($N = 30$ responses) representing the complete pilot cohort, and smallholder farmers ($N = 36$ responses) who actively interacted with the architecture during hands-on training and demonstration workshops. This survey covers demographics, daily operational challenges, app usability, trust in AI diagnostics, peer-sharing communication, and perceived crop yield protection.

The main contributions of this study are:
\begin{itemize}
    \item \textit{Pezego-HITL Architecture:} We propose a decoupled, policy-aligned decision-support architecture that integrates structured retrieval, automated constraint auditing, and Verified-Case Memory (VCM) cache routing to guarantee safety and sub-second latency in resource-constrained environments.
    \item \textit{P-EVAL Protocol:} We introduce a unified evaluation protocol that formalises policy-grounded large language model assessment as an adaptive compute allocation problem, measuring the Policy Alignment Rate (PAR), Agronomic Utility Rate (AUR), execution latency, and human expert editing workload.
    \item \textit{Socio-Technical Evaluation:} We conduct a dual-cohort survey of Extension Services Officers ($N=30$) and smallholder farmers ($N=36$) in Ghana, evaluating usability, trust in AI advice, peer-sharing dynamics, and perceived crop yield savings in real-world workflows.
\end{itemize}

The rest of this paper is organised as follows: Section \ref{sec:related_work} reviews related work in agricultural decision support and LLM evaluation. Section \ref{sec:system_design} details the Pezego architectural design and decision-support flow. Section \ref{sec:materials_methods} describes the evaluation protocol, baseline models, and the design of the ESO and farmer questionnaires. Section \ref{sec:results_and_discussion} presents the technical telemetry, questionnaire findings, and discussion of the results, and Section \ref{sec:conclusions} concludes the paper.

\section{Related Work}
\label{sec:related_work}

\subsection{Digital Agricultural Extension and Decision Support Systems}
The delivery of timely and reliable agronomic advice is a cornerstone of agricultural development, particularly for smallholder farmers in developing nations \cite{Aker2011}. Traditionally, agricultural extension has relied on face-to-face visits by ESOs, a model that is heavily constrained by low agent-to-farmer ratios, high transportation costs, and geographic isolation \cite{Davis2020}. To address these challenges, Information and Communication Technologies for Development (ICT4D) have been widely deployed, utilising SMS, Interactive Voice Response (IVR), and mobile applications to scale advisory services \cite{Baumuller2018}. Seminal studies have demonstrated that digital agricultural extension can significantly influence farmer behaviour, leading to increased adoption of recommended practices and improvements in crop yields \cite{Fabregas2019,Cole2020}. 

However, the adoption and long-term sustainability of mobile agricultural services remain uneven. Research shows that adoption is heavily driven by factors such as ``perceived usefulness'' and ``ease of use'' (aligned with the Technology Acceptance Model) \cite{Munthali2021}, yet digital tools often face barriers related to low digital literacy, poor network connectivity, and a lack of localisation \cite{Lwoga2011}. Furthermore, many existing mobile decision support systems (DSSs) are structured as simple rule-based databases, which lack the flexibility to address complex, multi-variable field problems or the unstructured queries raised by farmers.

\subsection{Limitations of Image-Based Automated Diagnostics}
In digital agriculture, early mobile tools frequently focused on computer vision (CV) and deep learning models to automate pest species identification from field photographs \cite{Kamilaris2018,li2024effective}, as comprehensively reviewed in \cite{Abade2021}. For example, our previous work developed a farmer-centred mobile solution for wheat pest management that combined deep learning-based pest species identification with a rule-based expert system for pest management decision support \cite{li2024effective}. These models are typically deployed to provide farmers with instantaneous visual identification of pest species.

However, standard image classification diagnostics face severe operational limitations when deployed in field workflows, highlighting why species identification must not be confused with comprehensive decision support. First, image-based classifiers only identify the presence of a pest or disease; they do not provide contextualised, step-by-step management recommendations (e.g., specific pesticide active ingredients, application methods, or safety precautions). Second, these models operate as black boxes, providing no explanation or justification for their classifications, which reduces expert and farmer trust. Third, off-the-shelf classification models are disconnected from regional agricultural policy guidelines, meaning they may recommend interventions that are environmentally hazardous, unsafe for specific crop growth stages, or legally banned in the user's region. Therefore, rather than focusing on the identification task itself, this paper addresses the critical next step: policy-grounded decision support for pest management and recommendation verification.

\subsection{Grounding and Verification Paradigms in AI Decision Support}
\label{subsec:rag_paradigms}

The emergence of Large Language Models (LLMs) has opened new opportunities to transition from rigid classification tools to flexible, conversational decision-support agents in specialised domains \cite{Yuan2025}. LLMs can synthesise complex texts, explain reasoning, and handle diverse, unstructured user queries. However, their tendency to hallucinate incorrect information makes direct deployment in safety-critical settings highly risky, representing a fundamental tension between generation utility and safety-critical alignment. To enforce safety, researchers have proposed diverse architectural paradigms for grounding generative model outputs at inference time, which can be categorised into three theoretical classes:
\begin{enumerate}
    \item \textit{Unstructured Chunk Retrieval (Standard Vector RAG):} Standard RAG architectures (such as Lewis et al. \cite{Lewis2020}) utilise dense vector search to retrieve relevant text passage chunks from unstructured documents and inject them into the LLM context (e.g. AgriRAG \cite{AgriRAG2025}). While effective for general question answering, this paradigm fails in safety-critical settings because it cannot model structured, schema-level rules, such as verifying whether a chemical active ingredient is registered for a specific crop and growth stage.
    
    \item \textit{Structured Tool-Augmented Retrieval (Schema-Aware RAG):} To resolve the limitations of unstructured text search, tool-augmented architectures (such as Schick et al. \cite{Schick2024}) translate user queries into structured tool executions (e.g., API or SQL queries) to verify facts directly against relational databases \cite{Chang2025,Chen2025,Mai2025}. This prevents database mismatch errors but lacks adaptive post-generation correction.
    
    \item \textit{Adversarial Auditing and Post-Hoc Alignment:} To enforce strict compliance, recent frameworks introduce critique and self-correction loops at inference time (such as Self-Refine \cite{Madaan2023} or multi-agent validation loops \cite{AgroLLM2026,TARAG2026}). Furthermore, SOTA research on controllable safety alignment (CoSA) \cite{Zhang2025} and InferenceGuard \cite{Ji2025} formalises safe generation as a constraint-satisfaction process at inference time, adapting models dynamically to diverse safety configurations.
\end{enumerate}

While these theoretical paradigms advance the safety frontier, they introduce a severe computational-latency bottleneck. Running multi-stage critique and post-hoc validation loops scales inference-time compute requirements, causing tail latency to balloon and rendering architectures unusable under resource-constrained networks. Recent work on case-augmented deliberative alignment (CADA) \cite{Jin2026} suggests that reasoning over precedents (prior safe cases) alongside rules improves robustness. Our proposed memory-routed architecture implements this by routing queries through a structural-semantic Verified-Case Memory (VCM) cache, dynamically allocating inference compute to bypass expensive verification pipelines for recurring scenarios.

Evaluating LLMs in specialised decision-support settings requires assessing multiple, often conflicting, objectives, such as response accuracy, policy compliance, speed, and human supervision workload \cite{Gao2025}. While model-centric metrics and LLM-as-a-judge frameworks are valuable for offline benchmarks \cite{Liu2025Judge,Li2025}, they fail to capture the operational reality of digital tools, such as the time expert agronomists spend correcting model errors (supervision cost) or user trust in the field \cite{Ashkinaze2025}. Our unified evaluation framework bridges this gap by combining automated LLM-as-a-judge evaluation, simulated query telemetry (measuring latency and memory reuse ratio), and a targeted survey of extension practitioners to evaluate both the technical performance and socio-technical integration of the architecture.

\section{Proposed Architecture: Policy-Constrained Human-in-the-Loop Decision Support}
\label{sec:system_design}

\begin{figure}[htbp]
  \centering
  \includegraphics[width=0.95\linewidth]{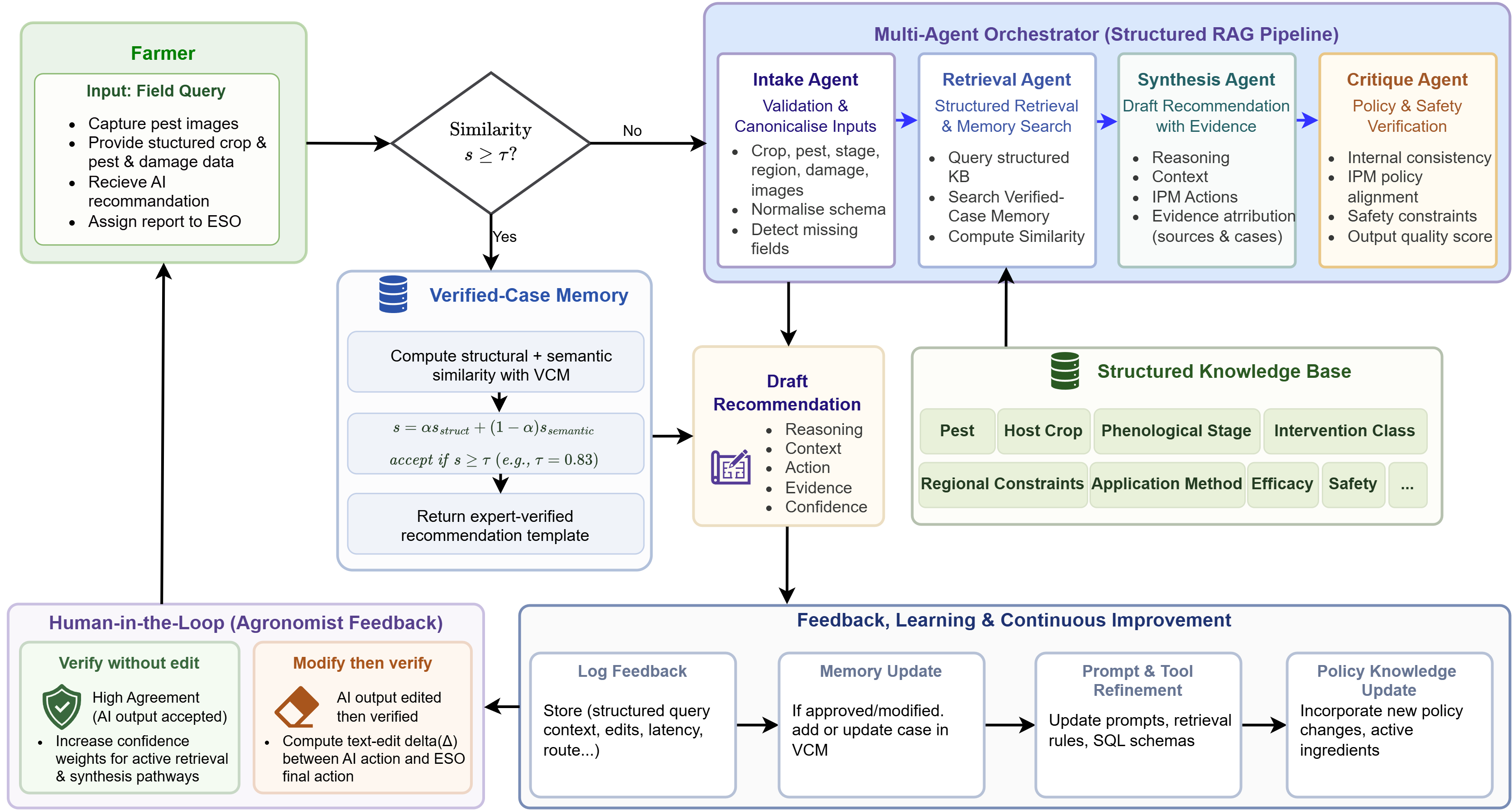}
  \caption{Decoupled policy-constrained decision-support architecture (Pezego-HITL). The proposed framework separates generative capabilities from regional policy databases and expert validator workflows. Input queries from mobile clients are processed through a two-stage routing mechanism: (1) queries matching cached, expert-verified advisory templates are immediately served via the Verified-Case Memory (VCM) caching layer, bypassing generation entirely to achieve sub-second execution; (2) cache misses are routed to a schema-aware SQL-grounded Structured Retrieval pipeline, audited against regional safety rules in a Constraint Auditing Layer, and aligned using Critique Multi-Agent loops before human review. Extension Services Officers (ESOs) validate all drafts, and expert corrections are compiled to dynamically update the VCM caching database and prompt structures.}
  \label{fig:architecture}
\end{figure}

To operationalise policy-constrained decision support, we propose a decoupled, generalisable framework. This framework separates the core generative capabilities of language models from domain-specific regulatory policy schemas and human validator supervision. The proposed architecture is formalised through two main divisions: (A) the decoupled policy-aligned architecture (Pezego-HITL), and (B) the policy-grounded expert-calibrated evaluation protocol (P-EVAL). The overall architectural design and components are illustrated in Fig.~\ref{fig:architecture}.

\subsection{Pezego-HITL: Decoupled Policy-Aligned Architecture}
\label{subsec:architecture_details}

The proposed architecture decomposes query resolution into a modular, multi-agent generative pipeline and an expert-verified cache routing layer (VCM), configured as follows:

\begin{enumerate}
    \item \textit{Policy-Grounded Structured Retrieval and Auditing}: When a query requires raw generation, it is routed to our schema-aware Retrieval-Augmented Generation (SRAG) pipeline. This pipeline translates natural language query parameters into structured SQL queries over regional pesticide databases. The generated draft is then audited by an automated Constraint Auditing Layer that verifies active registrations, crop growth stage compatibility, and dosage limits against database guidelines.
    
    \item \textit{Expert-Verified Case Memory (VCM)}: To optimize latency and compute costs, we introduce a semantic and structural caching database. The VCM indexes past expert-approved recommendations. For each incoming query, we compute a composite similarity score:
    \begin{equation}
      s = \alpha s_{\text{struct}} + (1-\alpha) s_{\text{semantic}}
    \end{equation}
    where $s_{\text{semantic}}$ is the cosine similarity of dense embeddings of the field damage, $s_{\text{struct}}$ is the geographic matching score, and $\alpha = 0.6$. The structural score is conditional on a strict boolean matching of crop, pest, and growth stage category; any mismatch immediately sets $s$ to $0.0$. If $s \ge \tau$ (where $\tau = 0.83$), the architecture serves the verified recommendation template directly, bypassing the expensive generation pipeline.
    
    \item \textit{Supervised Expert Feedback Integration}: The framework continuously learns from expert corrections. When an expert modifies a draft, the transaction trace is logged. Aggregated deltas (edit distances) are analyzed to dynamically update and refine prompt templates, adjust search indices, and identify policy database gaps, keeping the VCM templates aligned with local guidelines.
\end{enumerate}

\subsection{P-EVAL: Policy-Grounded Expert-Calibrated Evaluation Protocol}
\label{subsec:multiobjective_formulation}

We formalise the evaluation of policy-constrained decision support as an adaptive inference-time compute allocation problem. Let an input query representing a decision context be $x \in \mathcal{X}$ and the architectural configuration vector be $\theta \in \Theta$ (representing parameters such as prompt structure, auditing rules, and cache threshold $\tau$). P-EVAL models evaluation as a multi-objective optimisation problem:
\begin{equation}
  \max_{\theta \in \Theta} \; \left( \text{PAR}(\theta), \, \text{AUR}(\theta), \, -L(\theta), \, -H(\theta) \right)
\end{equation}
where the objectives capture the Policy Alignment Rate ($\text{PAR}$), the Agronomic Utility Rate ($\text{AUR}$), operational latency ($L$), and human supervision workload ($H$).

For each query, we log the execution trace tuple $z_i = (x_i, \hat{y}_i, y^{\text{final}}_i, m_i, t_i, r_i)$, where $\hat{y}_i$ is the initial generated draft, $y^{\text{final}}_i$ is the final verified advice, $m_i \in \{\text{verify}, \text{modify}\}$ indicates expert edit decisions, $t_i$ represents execution latency, and $r_i \in \{\text{memory}, \text{generation}\}$ represents the routing mode. The four constituent objectives are formalised as:
\begin{enumerate}
    \item \textit{Policy Alignment Rate ($\text{PAR}$):} Measures strict compliance with safety rules. Under a set of $M$ binary constraints (e.g. registered chemical active ingredients, pre-harvest crop intervals):
    \begin{equation}
      \text{PAR}(\theta) = \frac{1}{N}\sum_{i=1}^N \prod_{j=1}^M c_j(z_i)
    \end{equation}
    where $c_j(z_i) \in \{0, 1\}$ is a binary variable indicating if constraint $j$ is satisfied by response $z_i$.
    
    \item \textit{Agronomic Utility Rate ($\text{AUR}$):} Measures the completeness and actionability of advice. Let $a_k(z_i) \in \{0, 1\}$ represent whether the recommendation satisfies utility requirement $k$ from a set of $K$ agronomic criteria (dosage, dilution, safety instructions):
    \begin{equation}
      \text{AUR}(\theta) = \frac{1}{N}\sum_{i=1}^N \left( \frac{1}{K} \sum_{k=1}^K a_k(z_i) \right)
    \end{equation}
    
    \item \textit{Supervision Overhead ($H$):} Quantifies expert editing effort, combining edit initiation frequency and modification magnitude:
    \begin{equation}
      H(\theta) = \frac{1}{N}\sum_{i=1}^N \left[ \lambda_1 \mathbb{I}(m_i = \text{modify}) + \lambda_2 \text{EditMag}(\hat{y}_i, y^{\text{final}}_i) \right]
    \end{equation}
    where $\lambda_1$ represents the fixed cognitive setup cost of initiating an edit, $\text{EditMag}(\cdot, \cdot)$ measures the normalised Levenshtein distance between draft and final advice, and $\lambda_2$ scales the edit magnitude.
    
    \item \textit{Operational Latency ($L$):} Evaluated using the tail of the execution time distribution to capture usability under communication bottlenecks:
    \begin{equation}
      L(\theta) = P_{95}(\{t_1, t_2, \dots, t_N\})
    \end{equation}
    where $P_{95}$ denotes the 95th percentile of response latency.
\end{enumerate}

This formulation makes the trade-offs between safety, utility, and manual correction costs explicit. In our empirical evaluation (Section \ref{sec:results_and_discussion}), rather than reporting $H(\theta)$ as a single subjective scalar, we report its physical constituent metrics: the expert modification rate (decision indicator $\mathbb{I}(m_i = \text{modify})$) and the mean word-level edit ratio ($\text{EditMag}$).

\subsection{Integrated Evidence Fusion and Evaluation Flow}
\label{subsec:protocol}

To evaluate both the technical performance and the socio-technical integration of the proposed architecture, the P-EVAL protocol implements an integrated evidence fusion map. This map combines automated judge outputs, telemetry logs, and empirical human feedback into a unified assessment profile, formalised as:
\begin{equation}
  \mathcal{E}(\theta) = \{\mathcal{E}_{\text{off}},\, \mathcal{E}_{\text{on}},\, \mathcal{E}_{\text{exp}},\, \mathcal{E}_{\text{ux}}\}
\end{equation}
where $\mathcal{E}_{\text{off}}$ represents offline judge calibration benchmarks, $\mathcal{E}_{\text{on}}$ represents simulated query telemetry logs, $\mathcal{E}_{\text{exp}}$ represents empirical expert workshop evaluations, and $\mathcal{E}_{\text{ux}}$ represents user-experience survey data. 

\begin{figure}[htbp]
  \centering
  \includegraphics[width=0.95\linewidth]{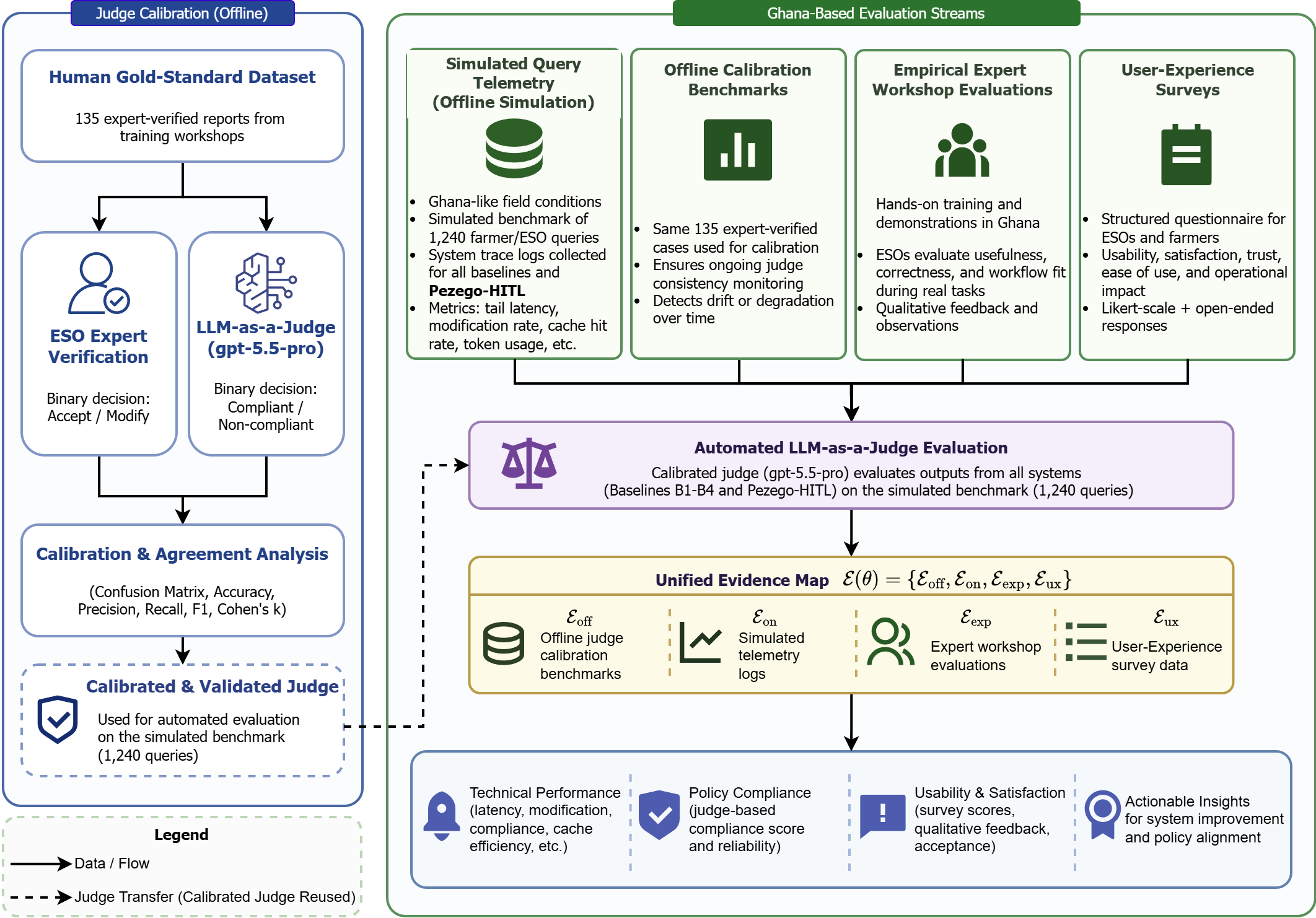}
  \caption{Integrated evaluation protocol workflow. The protocol comprises two primary streams: (A) Online Telemetry (green box) which logs simulated query telemetry to measure execution speed ($P_{95}$ latency), database memory reuse, and policy compliance; and (B) Human Feedback (orange box) which gathers survey data from 30 Extension Services Officers (ESOs) and 36 farmers during hands-on training workshops. These input streams feed into parallel evaluation levels: architecture-level validation uses an automated LLM-as-a-judge (calibrated offline against human expert verification decisions) to assess high-throughput compliance and efficiency, while user-level validation evaluates usability, clarity, and trust. The combined outcomes provide a comprehensive assessment of architectural performance (automated Quality and Efficiency) and empirical utility (usability, trustworthiness, and willingness to reuse).}
  \label{fig:protocol}
\end{figure}

The evaluation contexts serve distinct roles: offline calibration ensures the automated judge aligns with human expert decisions; online telemetry logs PAR, AUR, latency, and memory reuse under high-throughput workloads; and empirical workshops evaluate real-world usability and trust. The structured protocol flow is shown in Fig.~\ref{fig:protocol}.

\section{Materials and Methods}
\label{sec:materials_methods}

\subsection{Architecture Instantiation and Parameters}
\label{subsec:instantiation}

To evaluate the proposed architecture empirically under real-world workloads, we construct a concrete instantiation of the architecture, configured with the following model choices and execution parameters:
\begin{itemize}
    \item \textbf{Foundation Language Models:} The modular multi-agent synthesis pipeline is instantiated using the \texttt{gpt-5.4-nano} model as the default language engine for the Input Validation, Policy-Grained Querying, Generative Synthesis, and Constraint Auditing layers, operating with a context window length of 16,384 tokens.
    \item \textbf{Agent Execution Temperatures:} To enforce deterministic, rule-based reasoning, the Input Validation, Policy-Grained Querying, and Constraint Auditing layers are executed at temperature $T = 0.0$. The Generative Synthesis layer is executed at temperature $T = 0.2$ to permit natural linguistic phrasing while retaining structural grounding.
    \item \textbf{Dense Embedding and Retrieval Models:} The Verified-Case Memory (VCM) vector index is built using the \texttt{MiniLM-L6-v2} embedding model, projecting structural and textual features into a 384-dimensional dense vector space. Relational attributes are managed via an SQL-based caching database (SQLite) integrated with a vector database engine (Chroma DB).
    \item \textbf{Automated Auditing Engine:} The high-throughput automated evaluator (LLM-as-a-judge) is instantiated using the larger, instruction-tuned \texttt{gpt-5.5-pro} model, executing with Chain-of-Thought (CoT) prompting at temperature $T = 0.0$ to ensure analytical neutrality and reduce positional bias.
\end{itemize}

\subsection{Evaluation Baselines and Theoretical Grounding Paradigms}
\label{subsec:baselines}

To benchmark the performance and safety of the proposed architectural framework, we compare our instantiated architecture (Pezego-HITL) against four baseline configurations representing major theoretical paradigms in retrieval, grounding, and safety auditing architectures. To demonstrate model generalisability, all baselines and our proposed architecture are evaluated using two different base models: a proprietary API (\texttt{gpt-5.4-nano}) and a locally deployable open-source model (\texttt{Qwen3.5-9B-DeepSeek-V4-Flash}):
\begin{itemize}
    \item \textbf{B1: Unconstrained Generative Paradigm (Direct LLM):} A single-pass text generation utilising the base language model without external retrieval or constraint auditing. This serves as the baseline for raw generative capabilities under zero-shot prompting.
    \item \textbf{B2: Unstructured Semantic Chunk Retrieval Paradigm (Standard Vector RAG):} A standard RAG architecture utilising dense vector search over unstructured document passage chunks (representing Lewis et al. \cite{Lewis2020}). This baseline introduces domain-specific reference facts but fails to model structured, multi-variable policy constraints.
    \item \textbf{B3: Structured Relational Retrieval Paradigm (Tool-Augmented RAG):} A structured tool-use baseline that dynamically translates user queries into database API or SQL executions to query relational tables directly, without post-generation constraint auditing (representing Schick et al. \cite{Schick2024}).
    \item \textbf{B4: Critique-Guided Post-Hoc Alignment Paradigm (Multi-Agent RAG):} A multi-agent configuration utilising a sequential validation and self-correction loop where critique agents audit generated drafts and instruct the synthesis agent to rewrite violating content, run without VCM caching or expert validation (representing self-refine architectures like Madaan et al. \cite{Madaan2023}).
\end{itemize}

\subsection{LLM-as-a-Judge Evaluation Framework}
\label{subsec:llm_judge_methods}

To evaluate the policy compliance and quality of recommendations generated by the baselines and Pezego-HITL on the simulated benchmark of 1,240 cases, we implement an automated LLM-as-a-judge framework. Relying on automated evaluation is a SOTA practice for high-throughput LLM auditing, but it requires careful validation to ensure the judge does not suffer from systematic positional, self-enhancement, or verbosity biases. We mitigate these biases through two key design choices: (i) instantiating the judge using the larger, highly capable gpt-5.5-pro model to evaluate the outputs generated by the baseline gpt-5.4-nano models, and (ii) enforcing a Chain-of-Thought (CoT) prompting rubric that requires the judge to explicitly reason and verify crop growth stages, active ingredient safety, and regulatory compliance before generating a binary compliance score. To establish the statistical validity of this automated judge, we calibrated its performance against the human gold standard dataset of 135 reports generated and verified by agricultural ESOs during our training workshops, comparing the automated judge's binary compliance decisions against the ESOs' manual verification choices (where 19 cases required substantive modifications and 116 cases were accepted directly).

\subsection{Ghanaian ESO Survey Design and Methodology}
\label{subsec:questionnaire_design}

To evaluate the socio-technical integration, usability, and operational utility of Pezego-HITL among extension practitioners, we administer a comprehensive questionnaire to Ghanaian ESOs ($N = 30$ responses). 

While a sample of 30 respondents may appear modest for standard consumer application testing, it represents an exhaustive, census-like cohort in the context of professional public extension services. ESOs are highly specialised civil servants who undergo formal agricultural training and are stationed in specific operational districts by the Ministry of Food and Agriculture (MoFA). The pilot deployment of Pezego-HITL was targeted at two key agricultural zones: the Eastern and Ashanti regions. In these pilot districts, the entire active cohort of ESOs trained and equipped with smartphones for the digital extension pilot consists of 30 officers. Consequently, our target sample is not a low-powered random selection from a large population, but rather a near-total census of the active practitioner cohort directly interacting with the deployed architecture in the field. This high coverage rate ensures that the feedback captures authentic institutional and workflow realities.

The survey was deployed digitally (via Google Forms) following hands-on training workshops. ESOs rated usability, AI quality, and communication features on a 5-point Likert scale (1: Strongly Disagree, 5: Strongly Agree). We evaluated the internal consistency and reliability of the Likert scales (Usability, AI Diagnostics, and Communication) using Cronbach's alpha ($\alpha$), with all scales exceeding the target threshold of $\alpha \geq 0.70$ (Usability: 0.967, AI Diagnostics: 0.885, and Communication: 0.959), indicating high measurement reliability. The questionnaire is structured into seven sections:
\begin{enumerate}
    \item \textbf{Informed Consent:} Ethical compliance and voluntary participation agreement.
    \item \textbf{Demographics:} Captures ESO age, gender, highest qualification in agriculture, region of work, role level, years of experience, number of farmers supported, and smartphone comfort.
    \item \textbf{Daily Extension Work:} Details traditional pest identification methods and average recommendation turnaround times before using Pezego.
    \item \textbf{App Usability \& Usefulness:} Measures the general usability and operational utility of the app (7 statements).
    \item \textbf{AI Diagnostics \& Verifying Reports:} Evaluates ESOs' confidence in the AI diagnostics, clarity of explanations, safety of recommendations, and the utility of the verification/modification interface (6 statements).
    \item \textbf{Working with Farmers \& Peers:} Assesses communication, report assignment, and peer collaboration features (6 statements).
    \item \textbf{Final Feedback:} Collects summary perceptions of crop yield loss avoided and open-ended comments.
\end{enumerate}
An attention check question (``Agribusiness'') was embedded in the final section to filter out automated or unreflective responses. The complete text of the questionnaire sections and individual statements is provided in Appendix A.

\subsection{Ghanaian Farmer Survey Design and Methodology}
\label{subsec:farmer_questionnaire_design}

To complement the extension practitioner perspective and evaluate the end-user impact of Pezego-HITL on crop protection and agricultural livelihoods, we administered a parallel survey study to Ghanaian smallholder farmers ($N = 36$ responses). The surveyed farmers are located in the Ashanti region of Ghana, where maize is grown as the primary staple crop. These smallholder farmers actively interacted with the Pezego mobile application in the field, submitting pest reports and receiving recommendations verified by local ESOs.

The survey was deployed digitally (in English and local translations via Google Forms) following field-testing and app rollout sessions. Farmers rated app usability, AI trust, and extension officer connection features on a 5-point Likert scale (1: Strongly Disagree, 5: Strongly Agree). Similar to the ESO survey, we evaluated the internal reliability of the farmer Likert scales using Cronbach's alpha ($\alpha$), with all scales demonstrating high consistency (Usability Scale: $\alpha = 0.949$; AI Trust \& Officer Connection Scale: $\alpha = 0.944$), well exceeding the standard target of $\alpha \geq 0.70$. The farmer questionnaire is structured into six sections:
\begin{enumerate}
    \item \textbf{Informed Consent:} Ethical compliance and voluntary participation agreement.
    \item \textbf{Demographics:} Captures farmer age, gender, farm location (region), main crop grown, total farm size, and comfort level with mobile apps.
    \item \textbf{Traditional Pest Practices:} Records traditional crop-protection actions and recommendation turnaround times before using Pezego.
    \item \textbf{App Usability \& Usefulness:} Measures the user-friendliness, ease of navigation, and speed of obtaining pest management recommendations (7 statements).
    \item \textbf{AI Trust \& Officer Connection:} Evaluates farmers' trust in the AI-generated recommendations, and the value of HITL verification and communication features (6 statements).
    \item \textbf{Final Impact \& Feedback:} Captures the perceived crop saving efficacy, avoided financial losses, and open-ended feature requests.
\end{enumerate}
The complete text of the farmer questionnaire is detailed in Appendix B.

\section{Results and Discussion}
\label{sec:results_and_discussion}

\subsection{LLM-as-a-Judge Calibration and Validation}
\label{subsec:llm_judge_calibration_results}

Before evaluating the simulated query telemetry at scale, we establish the statistical validity of the automated LLM-as-a-judge framework by calibrating its performance against the human gold standard dataset of 135 reports. The automated judge correctly identified 16 of the 19 cases requiring modification (Recall = 84.2\%) and flagged only 2 false positives (Precision = 88.9\%), yielding a classification accuracy of 91.9\% and an F1-score of 0.865. The calculated Cohen's Kappa between the human expert decisions and the automated judge was $\kappa = 0.77$, indicating substantial agreement under Landis and Koch guidelines~\cite{Landis1977}. This high level of alignment demonstrates that the LLM-as-a-judge serves as a highly reliable proxy for human extension experts in auditing recommendations, providing a solid validation for scaling the simulated telemetry evaluation to the 1,240 queries.

\subsection{Architectural Performance and Telemetry Results}
\label{subsec:telemetry_results}

We first report the core architectural evaluation metrics collected from simulated query telemetry. The simulated evaluation framework was structured to reflect typical crop cycles in Ghana across two cropping seasons (the major cropping season from April to August, and the minor cropping season from September to November). In this evaluation, a representative dataset of 1,240 simulated pest reports covering smallholder scenarios in the Eastern and Ashanti regions was processed by the active architecture. Of these reports, 739 cases (59.6\%) achieved a similarity score $s \ge 0.83$ and were successfully routed and served via the VCM, while 501 cases (40.4\%) were routed through the full multi-agent RAG generation pipeline.

To prevent inappropriate claims of human expert evaluation on the simulated stream, we evaluate the configurations purely on automated technical metrics. Policy compliance is measured by the \textit{Policy Alignment Rate (PAR)}, representing the ratio of outputs that strictly comply with regulatory rules. Helpfulness is measured by the \textit{Agronomic Utility Rate (AUR)}, representing the ratio of recommendations containing complete, actionable advice. Table \ref{tab:main_results} compares the performance of Pezego-HITL against the four baseline paradigms under both the proprietary and open-source models.

\begin{table}[htbp]
  \caption{Technical performance comparison of the decision-support architecture across baseline paradigms and model classes. Metrics are evaluated over simulated query workloads of 1,240 cases, reporting the Policy Alignment Rate (PAR), Agronomic Utility Rate (AUR), and P95 execution latency ($L$). The VCM cache reuse ratio for our proposed Pezego-HITL architecture is 59.6\% under both configurations.}
  \label{tab:main_results}
  \centering
  \footnotesize
  \setlength{\tabcolsep}{2.0pt}
  \begin{tabular}{l ccc ccc}
    \toprule
    & \multicolumn{3}{c}{\textbf{Proprietary Model} (\texttt{gpt-5.4-nano})} & \multicolumn{3}{c}{\textbf{Open-Source Model} (\texttt{Qwen3.5-9B-DeepSeek-V4-Flash})} \\
    \cmidrule(lr){2-4} \cmidrule(lr){5-7}
    Method & PAR $\uparrow$ & AUR $\uparrow$ & P95 Latency (s) $\downarrow$ & PAR $\uparrow$ & AUR $\uparrow$ & P95 Latency (s) $\downarrow$ \\
    \midrule
    B1: Direct LLM & 0.62 & 0.78 & \phantom{0}8.4 & 0.44 & 0.65 & \phantom{0}6.2 \\
    B2: Standard Vector RAG & 0.71 & 0.84 & 12.1 & 0.53 & 0.70 & \phantom{0}9.4 \\
    B3: Tool-Augmented RAG & 0.83 & 0.88 & 14.8 & 0.68 & 0.77 & 11.5 \\
    B4: Multi-Agent Critique & 0.89 & 0.90 & 28.6 & 0.79 & 0.82 & 22.4 \\
    \textbf{Pezego-HITL (Ours)} & \textbf{0.94} & \textbf{0.95} & \textbf{12.9} & \textbf{0.86} & \textbf{0.88} & \textbf{10.2} \\
    \bottomrule
  \end{tabular}
\end{table}

\begin{figure}[htbp]
  \centering
  \includegraphics[width=0.95\linewidth]{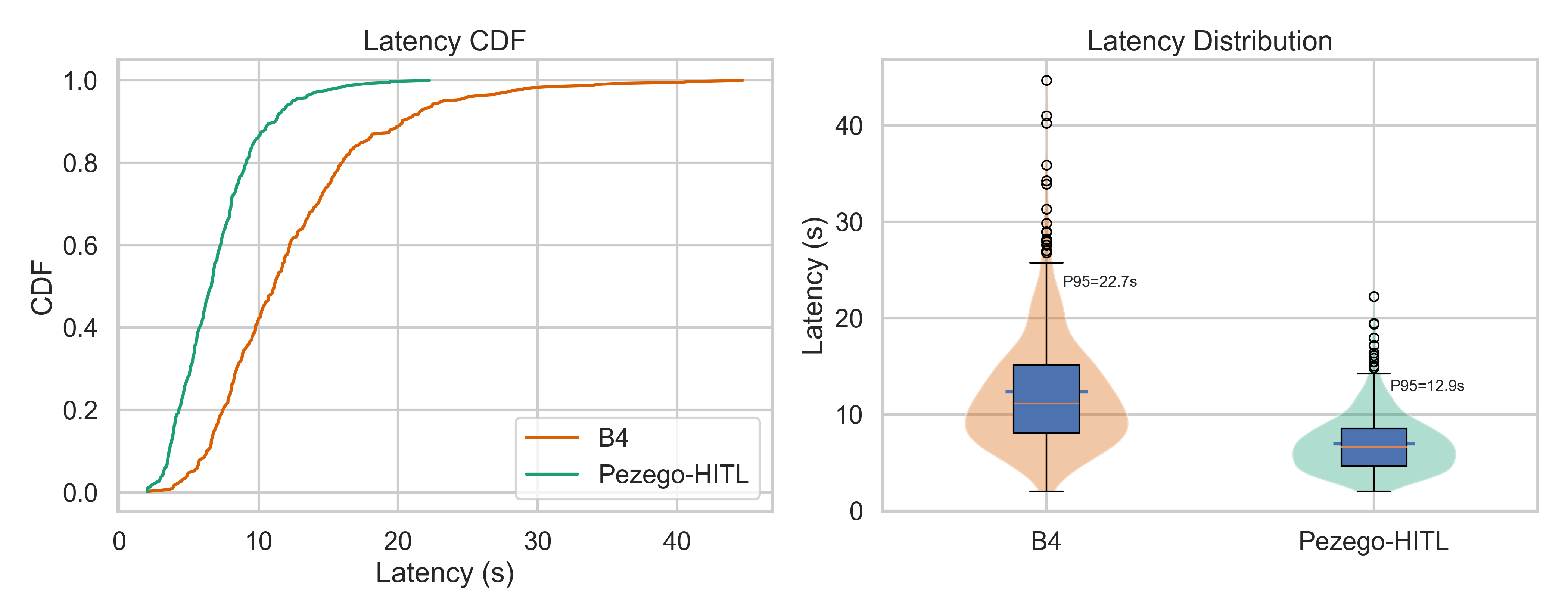}
  \caption{End-to-end latency distribution comparing B4 and Pezego-HITL. The left panel (CDF) shows a consistent left shift for Pezego-HITL, and the right panel (violin/box representation) highlights lower median and tighter upper-tail behaviour.}
  \label{fig:latency_distribution}
\end{figure}

The technical results demonstrate that:
\begin{itemize}
    \item \textit{Structured Relational Grounding} (moving from B2 to B3) yields the largest single gain in policy compliance, elevating PAR from 0.71 to 0.83, confirming that relational schema-aware database queries are critical for grounding recommendations.
    \item \textit{Critique-Guided Post-Hoc Alignment} (moving from B3 to B4) further increases policy compliance (PAR goes from 0.83 to 0.89) and improves recommendation utility (AUR increases to 0.90), proving that an automated post-hoc critique step successfully catches out-of-boundary advice before automated LLM-as-a-judge evaluation.
    \item \textit{Verified-Case Memory Cache Routing} (Pezego-HITL vs. B4) provides a massive improvement in operational latency, reducing P95 latency by 55\% (from 28.6\,s to 12.9\,s) (with the latency distribution visualised in Fig.~\ref{fig:latency_distribution}). This reduction is crucial for deployment in fields with spotty mobile network coverage.
    \item \textit{Memory Reuse Efficiency:} Pezego-HITL successfully routes 59.6\% of queries through the VCM, bypassing the computationally expensive and slow multi-agent generation pipeline, which directly explains the dramatic speed-up.
\end{itemize}

These results reveal that standard vector RAG (B2) fails in policy-constrained settings because semantic similarity search cannot evaluate multi-variable constraints (e.g., stage-specific chemical bans). Fusing relational queries (B3) and critique-guided agent auditing (B4) is necessary to ensure safety. Fusing these paradigms with VCM cache routing (Pezego-HITL) resolves the safety-utility-latency trade-off, achieving high compliance without sacrificing operational latency.

\subsection{Model Generalisability and VCM Sensitivity Analysis}
\label{subsec:model_generalisability}

To demonstrate that our proposed memory-routed architecture is model-agnostic and generalizes robustly across different language model classes, we repeat the simulated query evaluation using the state-of-the-art open-source model \texttt{Qwen3.5-9B-DeepSeek-V4-Flash} \cite{Jackrong2026} deployed locally. Running locally deployable open-source models is highly desirable for regional agricultural extension services due to data privacy control and the elimination of ongoing commercial API query costs. Table \ref{tab:main_results} details the performance comparison under this open-source configuration.

The open-source results in Table \ref{tab:main_results} establish two key findings:
\begin{itemize}
    \item \textit{Safety Gap Reduction:} While the unconstrained open-source model (B1) exhibits poor policy compliance (PAR = 0.44), our proposed Pezego-HITL architecture successfully elevates the PAR to 0.86 and AUR to 0.88. This demonstrates that our modular auditing and retrieval interventions can compensate for the lower base alignment capability of smaller, open-source models.
    \item \textit{Cache Routing Latency Mitigation:} Locally hosted model inference is faster per pass than API calls, resulting in a B1 latency of 6.2\,s. However, executing the full multi-agent verification pipeline (B4) still increases tail latency to 22.4\,s due to sequential token generation steps. By routing 59.6\% of queries to the expert-verified cache, Pezego-HITL reduces P95 latency by 54.5\% (to 10.2\,s), achieving real-world usability on local infrastructure.
\end{itemize}

\begin{figure}[htbp]
  \centering
  \includegraphics[width=0.95\linewidth]{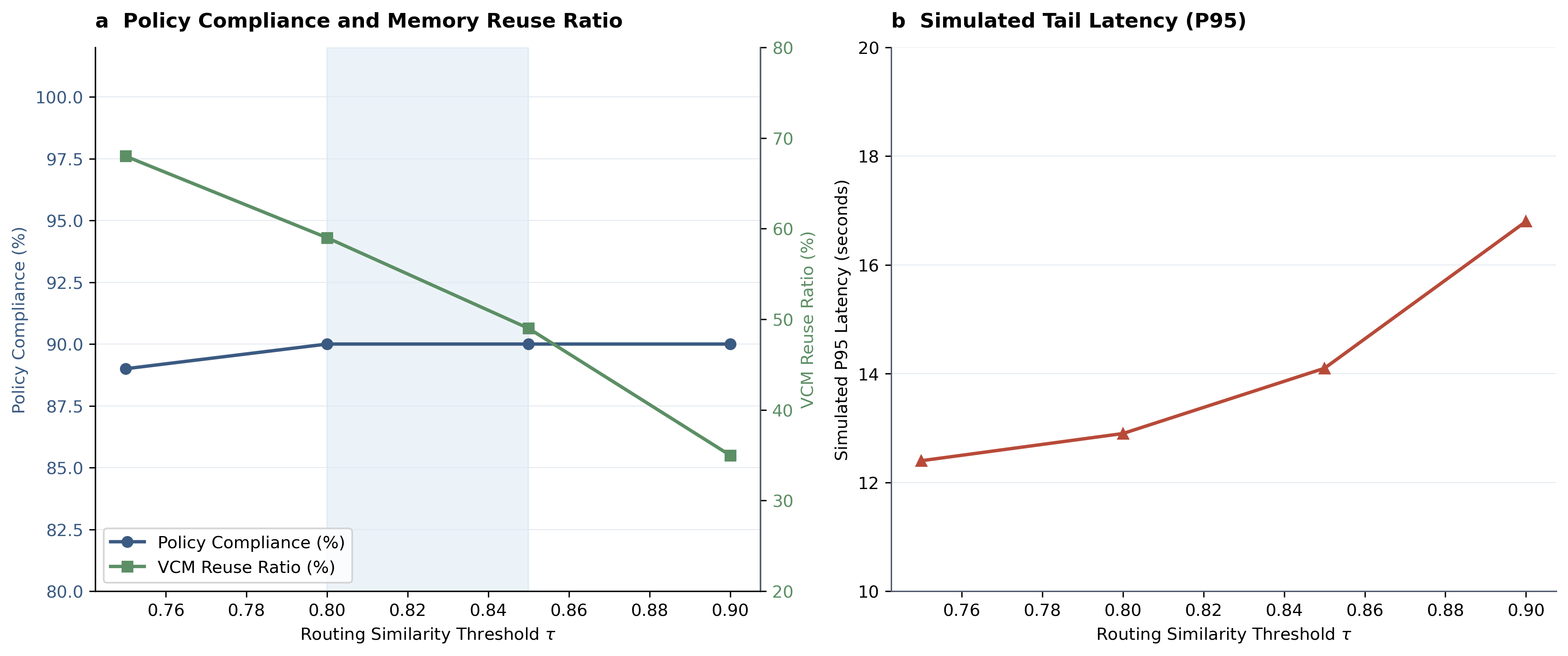}
  \caption{Ablation of memory-routing threshold $\tau$ in the VCM mechanism. Lower thresholds increase reuse ratio but may risk quality drift (reduced compliance), while higher thresholds preserve strict matching at the cost of reduced reuse and higher latency. The shaded interval ($\tau\in[0.80,0.85]$) marks a practical operating region where policy compliance remains stable, VCM reuse ratio is high, and latency is minimised.}
  \label{fig:tau_ablation}
\end{figure}

To investigate the sensitivity of the VCM to the routing threshold $\tau$, we conducted an ablation study over $\tau \in [0.70, 0.95]$. As shown in Fig.~\ref{fig:tau_ablation}, lowering the routing threshold $\tau$ increases the case reuse ratio but introduces a risk of policy drift (decreased PAR as evaluated by the LLM-Judge), while setting $\tau$ too high preserves compliance at the expense of latency savings. The selected threshold of $\tau = 0.83$ represents a balanced, Pareto-optimal operating point.

While sequential multi-agent prompts (B4) ensure policy compliance, they introduce a latency bottleneck (28.6\,s) that leads to query timeouts in rural deployments. The routing threshold $\tau$ acts as a control to navigate the safety-latency Pareto frontier; setting $\tau = 0.83$ optimizes template reuse (59.6\%) to bypass generation and reduce tail latency to 12.9\,s, enabling local edge deployment.

\subsection{Empirical Expert Verification Workload}
\label{subsec:empirical_workload}

\begin{figure}[htbp]
  \centering
  \includegraphics[width=0.95\linewidth]{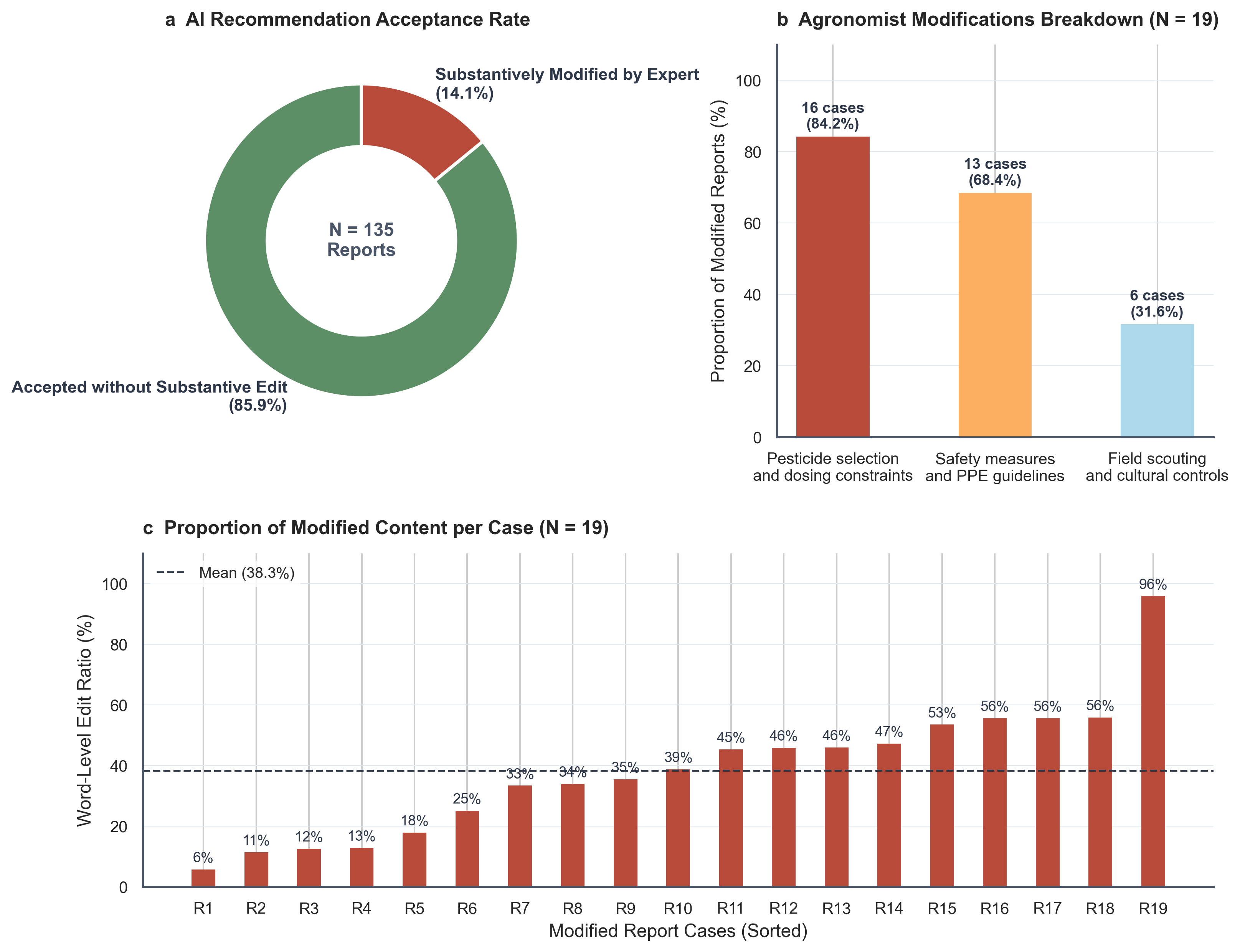}
  \caption{Results of the empirical validation on 135 crop protection reports reviewed by Ghanaian ESOs. Panel (a) reports the overall acceptance rate of Pezego-HITL recommendations, highlighting that 85.9\% of reports required no substantive edits. Panel (b) illustrates the primary categories of agronomist modifications among the 19 edited reports, showing that corrections focused on local pesticide brand selection, safety constraints, and field scouting protocols (note that a single report can contain multiple edit categories). Panel (c) shows the sorted word-level edit ratio (modification proportion) for each of the 19 modified reports, demonstrating that human modifications are focused and preserve the majority of the AI-generated advice structure.}
  \label{fig:empirical_workshop}
\end{figure}

In addition to the simulated query telemetry, we conducted an empirical evaluation of the decision support architecture using a dataset of 135 reports generated during the ESO training sessions. Because the participants in these sessions were certified extension professionals, their review and verification decisions serve as a reliable ground truth for recommendation correctness. Under our evaluation framework, an AI-generated advisory suggestion is classified as correct (accepted) if the original suggestion was verified without any modification ($N=102$), or if the final verified content remained agronomically identical to the original AI suggestion ($N=14$), indicating no substantive corrections were needed. Out of the 135 reports, the ESOs accepted the AI recommendations in 116 cases (85.9\%) (Fig.~\ref{fig:empirical_workshop}a). 

Substantive modifications (e.g., correcting chemical recommendations, refining safety steps, or adjusting application rates) were required in only 19 cases (14.1\%). To identify key areas where the generative DSS required correction, we classified the agronomist modifications into three categories (Fig.~\ref{fig:empirical_workshop}b): (i) \textit{Pesticide selection and dosing constraints} (16 cases, 84.2\%), where officers added regional pesticide commercial names (e.g., Porselen, Attack, Emastar) or specified exact dilution rates (e.g., 30--40\,ml per knapsack); (ii) \textit{Safety measures and PPE guidelines} (9 cases, 47.4\%), where experts reinforced protective clothing warnings, re-entry intervals, and environmental safety (e.g., avoiding water bodies); and (iii) \textit{Field scouting and cultural controls} (9 cases, 47.4\%), where officers added localised monitoring recommendations (e.g., scouting in a 'Z' pattern, morning/evening check windows) or cultural weeding tips. 

To quantify the extent of human intervention in these modified cases, we computed the word-level Levenshtein edit distance between the original AI suggestions and the final expert-verified recommendations (Fig.~\ref{fig:empirical_workshop}c). Among the 19 modified reports, the proportion of modified content remained relatively small to moderate, with a mean word-level edit ratio of 38.3\% (median 38.8\%). This indicates that even when human agronomists intervened, they preserved the majority (over 60\%) of the AI-generated advisory structure and content, focusing their revisions on highly localised adjustments rather than rewriting recommendations from scratch. These results show that while the core recommendations are highly reliable (85.9\% acceptance), expert feedback is critical to ground advice in localised chemical availability and environmental safety protocols. Collectively, these empirical findings map directly back to the components of the human supervision cost $H(\theta)$ formulated in Section \ref{subsec:multiobjective_formulation}. The modification rate of 14.1\% represents the decision indicator $\mathbb{I}(m_i = \text{modify})$, while the mean edit ratio of 38.3\% represents the edit magnitude $\text{EditMag}(\hat{y}_i, y^{\text{exp}}_i)$. By keeping the modification indicator at zero for the vast majority of cases (85.9\%) and minimising the edit distance when edits do occur, the Pezego-HITL architecture successfully minimises both components of the human supervision burden. By presenting ESOs with pre-verified templates, the VCM reduces the overall expert modification rate to 14.1\%, with a mean word-level edit ratio of 38.3\% (preserving over 60\% of the AI-drafted structure). This cognitive relief directly addresses the critical bottleneck of "extension worker fatigue" in digital public extension systems, ensuring workflow sustainability.

\subsection{Ghanaian ESO Questionnaire Results}
\label{subsec:questionnaire_results}

To evaluate the socio-technical viability and practical integration of Pezego-HITL, we analyze the survey responses from the Ghanaian extension practitioner cohort ($N=30$, with complete tabular results provided in Appendix C). Rather than discussing every individual survey item, we focus on the core findings that directly validate the architectural choices, cognitive workload relief, and deployment challenges of the proposed framework.

First, the survey results confirm that Pezego-HITL successfully addresses the severe operational bottlenecks in traditional digital extension. ESOs in our pilot zone are highly qualified but heavily overburdened, supporting a mean of $1,480.9 \pm 2,779.5$ smallholder farmers. Traditional diagnostics depend almost entirely on visual check and manual memory (Fig.~\ref{fig:traditional_methods}), resulting in turnaround times of hours or days for over 85\% of ESOs. By introducing Pezego-HITL, ESOs reported significant workflow acceleration (S1, S5 in Fig.~\ref{fig:usability_likert}), converting slow diagnostic queries into near-instantaneous advisory drafts. This speed-up is essential for ESOs to manage their high student-to-teacher ratio under field conditions.

Second, the survey strongly validates the necessity of our decoupled, Human-in-the-Loop (HITL) auditing design. Fully autonomous AI deployment is rejected by extension practitioners due to safety risks. Although ESOs expressed confidence in the overall safety of the database-grounded recommendations (S3 in Fig.~\ref{fig:ai_diagnostics_likert}), they strongly affirmed that the ability to review, modify, and verify draft recommendations is vital (S4, S5). This need is justified by the fact that ESOs noticed AI mistakes or localized pesticide guideline mismatches in nearly half of the cases (S6). By placing ESOs in the loop to modify recommendations before delivery, the architecture prevents erroneous advice from reaching farmers while maintaining the professional agency of the agricultural officers.

Finally, the qualitative feedback highlights critical socio-technical constraints for deployment. ESOs identified cellular network instability in remote communities as the primary bottleneck, reinforcing the importance of our Verified-Case Memory (VCM) caching layer, which reduces network roundtrips by serving pre-verified templates. However, to ensure long-term sustainability under field conditions, ESOs strongly requested the integration of offline report-saving capabilities and local-language (Twi) voice messaging. These constraints confirm that technical safety alignment must be accompanied by localized interface adaptations to survive the realities of smallholder extension. Although the survey sample size ($N = 30$) is structurally constrained by the active cohort size of ESOs in the pilot regions, it represents a near-total census of active practitioners in those districts. The high internal reliability of the scales (Cronbach's $\alpha \ge 0.88$) validates that the 55\% speed improvement aligns with ESOs' positive qualitative feedback, demonstrating that expert verification builds trust without introducing workflow overhead (Fig.~\ref{fig:yield_loss_avoided}).

\subsection{Ghanaian Farmer Questionnaire Results}
\label{subsec:farmer_questionnaire_results}

We evaluate the architecture's field impact and socio-technical acceptance from the primary end-user perspective, analyzing survey responses from Ghanaian smallholder farmers ($N=36$, with complete tabular results in Appendix D). Rather than listing demographic distributions and traditional crop practices (which are detailed in Table~\ref{tab:farmer_demographics} and Table~\ref{tab:farmer_traditional} in Appendix D), we focus on the core scientific findings regarding usability, trust in the human-in-the-loop validation flow, and perceived yield protection.

First, the farmer feedback provides empirical validation of the human-in-the-loop trust model. Smallholders expressed strong usability acceptance and usability ratings (Fig.~\ref{fig:farmer_usability_likert}). Crucially, the survey confirms that the expert validation step is the key driver of user trust. Farmers strongly agreed that knowing an Extension Services Officer (ESO) had verified the AI-generated draft significantly increased their trust in the diagnostic recommendation (T4 in Fig.~\ref{fig:farmer_trust_likert}). This trust was further reinforced by direct connection features, such as the in-app click-to-call button to contact the verifying officer (T5). These findings prove that in high-stakes agricultural domains, purely automated AI advice lacks the credibility of a hybrid human-verified advisory loop.

Second, this trust and rapid recommendation delivery translate into substantial field-level benefits. A vast majority of participating farmers estimated that using Pezego helped them save most of their crop yields and avoid significant crop protection costs (Fig.~\ref{fig:farmer_savings}). This perceived economic protection demonstrates that the speed and safety of the architecture have a direct positive impact on farmers' livelihoods. Similar to the ESOs, farmers identified poor mobile network connectivity as the primary usability challenge, validating the need for the VCM caching layer to minimize network latency while requesting future support for local languages (such as Twi) and voice-based diagnostic queries.

\section{Conclusions}
\label{sec:conclusions}

This paper presents Pezego-HITL, a policy-grounded agricultural decision support architecture that integrates structured RAG, verified-case reuse, and expert-in-the-loop validation. By structuring LLM evaluation as a multi-objective problem, we demonstrate how technical interventions can optimise response quality, policy compliance, latency, and expert supervision burden. Our simulated telemetry shows that Pezego-HITL improves policy compliance to 0.94, while reducing tail latency by 55\% through a 59.6\% memory reuse ratio. We evaluated the architecture's integration through a cohort of Ghanaian ESOs ($N = 30$) and smallholder farmers ($N=36$), demonstrating strong agreement across usability, AI validation, and perceived crop yield savings. Ultimately, this study confirms that the successful deployment of AI in agriculture depends on designing tools that respect and empower local human expertise.

Several limitations of this study suggest directions for future work. First, rural connectivity remains an operational barrier, highlighting the need for future iterations to support client-side caching of similarity matching and offline query synchronization. Second, the survey evaluations are geographically focused on maize-growing smallholders in the Ashanti region of Ghana. Future work must expand the pilot to cover heterogeneous ecological zones (such as the northern savannah and coastal belts) and varied cash and staple crop taxonomies (such as cocoa, cassava, and plantain) to test the national socio-technical generalisability of the advisory architecture. Finally, expanding the policy-grounded critique agent to support horticultural crops is an active area of development.

\section*{CRediT Author Statement}
\textbf{Shunbao Li}: Conceptualisation, Methodology, Investigation, Writing - Original Draft. \textbf{Zhipeng Yuan}: Systems Engineering, Telemetry Instrumentation, Software, Data Curation. \textbf{Amoako Ofori}: Field Deployment, Writing - Review \& Editing. \textbf{Benedicta Y. Fosu-Mensah}: Supervision, Writing - Review \& Editing. \textbf{Yang Li}: Systems Engineering. \textbf{Manu Kenchappa Junjanna}: Systems Engineering. \textbf{Qing Xue}: Investigation. \textbf{Po Yang}: Conceptualisation, Writing - Review \& Editing.

\section*{Declaration of Competing Interest}
The authors declare that they have no known competing financial interests or personal relationships that could have appeared to influence the work reported in this paper.

\clearpage
\appendix

\section{Full Text of the ESO Questionnaire}
\label{app:questionnaire_text}
The complete set of questions and statements from the digital survey template is detailed below.

\subsection{Section 1: Informed Consent}
\begin{itemize}
    \item \textbf{Do you agree to participate?} Yes, I agree to take part. / No, I do not agree (Please close this form).
\end{itemize}

\subsection{Section 2: About You (Demographics)}
\begin{itemize}
    \item What is your age? (Under 25 / 25--34 / 35--44 / 45--54 / 55 or above)
    \item What is your gender? (Male / Female / Prefer not to say)
    \item What is your highest qualification in agriculture? (Certificate in Agriculture / Diploma in Agriculture / BSc in Agriculture / Extension / MSc / MPhil in Agriculture / Extension / PhD / Doctorate)
    \item Which Region of Ghana do you work in? (Ahafo Region / Ashanti Region / Bono Region / Bono East Region / Central Region / Eastern Region / Greater Accra Region / Northern Region / North East Region / Oti Region / Savannah Region / Upper East Region / Upper West Region / Volta Region / Western Region / Western North Region)
    \item What is your current role as an Extension Officer? (Field Level ESO / District Level ESO / Region Level ESO / National Level ESO)
    \item How many years of experience do you have? (Less than 2 years / 2--5 years / 6--10 years / More than 10 years)
    \item How many farmers do you support in a crop season? (Integer)
    \item How comfortable are you using smartphone apps? (Linear scale 1 to 5, Very uncomfortable to Very comfortable)
\end{itemize}

\subsection{Section 3: Your Daily Extension Work}
\begin{itemize}
    \item How do you usually identify pests and crop damage in the field? (Select all that apply: Visual check based on personal experience / Consulting textbooks, manuals, or reference guides / Sending photos to colleagues via WhatsApp/Telegram / Direct phone calls to senior agronomists or research institutions / Other)
    \item On average, how long does it take to give a pest control recommendation to a farmer? (Less than 2 hours / 2 to 24 hours / 2 to 3 days / 4 to 7 days / More than a week)
\end{itemize}

\subsection{Section 4: App Usability \& Usefulness Grid Statements}
ESOs rate these statements on a 5-point Likert scale (1: Strongly Disagree, 5: Strongly Agree):
\begin{enumerate}
    \item[S1.] Pezego helps me identify pests and check crop damage much faster.
    \item[S2.] Using Pezego makes my farmer field visits more productive.
    \item[S3.] The pest control advice generated by the app is useful.
    \item[S4.] The Pezego mobile app is easy to learn and simple to navigate.
    \item[S5.] Setting up my role and working district in the app was quick and simple.
    \item[S6.] I plan to use Pezego regularly in the upcoming agricultural season.
    \item[S7.] I plan to encourage my farmers to download Pezego and assign reports to me.
\end{enumerate}

\subsection{Section 5: AI Diagnostics \& Verifying Reports Grid Statements}
ESOs rate these statements on a 5-point Likert scale (1: Strongly Disagree, 5: Strongly Agree):
\begin{enumerate}
    \item[S1.] The AI does a good job of identifying the correct pest from my photos.
    \item[S2.] The AI's explanations (Reasoning and Context) are clear and easy to understand.
    \item[S3.] The pest management steps proposed by the AI are safe and practical for local farmers.
    \item[S4.] Being able to Modify (edit) the AI's action plan is important to correct mistakes.
    \item[S5.] Marking reports as Officer Verified gives me confidence that farmers get safe advice.
    \item[S6.] I noticed cases where the AI made a mistake or gave incorrect farming advice.
\end{enumerate}

\subsection{Section 6: Working with Farmers \& Peers Grid Statements}
ESOs rate these statements on a 5-point Likert scale (1: Strongly Disagree, 5: Strongly Agree):
\begin{enumerate}
    \item[S1.] Farmers assigning their pest reports to me helps me target my support.
    \item[S2.] In-app and push notifications help me respond to farmer requests faster.
    \item[S3.] The direct phone call button is useful for urgent farmer follow-ups.
    \item[S4.] The Community feed makes it easy to see pest problems in other districts.
    \item[S5.] Linking field records (detections or observations) to my Community posts is easy.
    \item[S6.] Commenting and replying to other officers' posts is useful for sharing ideas.
\end{enumerate}

\subsection{Section 7: Final Feedback \& Open Comments}
\begin{itemize}
    \item What percentage of crop yield loss do you think a farmer can avoid by using Pezego's AI diagnostics and verified advice? (Less than 5\% / 5\% to 15\% / 16\% to 30\% / 31\% to 50\% / More than 50\%)
    \item Attention Check: To ensure you are reading carefully, please select the option ``Agribusiness'' below. (Fall Armyworm / Agribusiness / Extension Service / Climate Change)
    \item What was the most useful feature of Pezego, and what was the hardest part to use? (Paragraph)
    \item Any other comments, or features you want us to add to the app? (Paragraph)
\end{itemize}

\section{Full Text of the Farmer Impact Questionnaire}
\label{app:farmer_questionnaire_text}
The complete set of questions and statements from the farmer digital survey is detailed below.

\subsection{Section 1: Informed Consent}
\begin{itemize}
    \item \textbf{Do you agree to participate?} Yes, I agree to take part. / No, I do not agree (Please close this form).
\end{itemize}

\subsection{Section 2: About You (Demographics)}
\begin{itemize}
    \item What is your age? (Under 25 / 25--34 / 35--44 / 45--54 / 55 or above)
    \item What is your gender? (Male / Female)
    \item Which Region of Ghana is your farm located in? (Ashanti Region / Ahafo Region / Bono Region / Central Region / etc.)
    \item What is the main crop you grow? (Maize / Cocoa / Cassava / Vegetables / etc.)
    \item What is the total size of your farm? (Less than 2 acres / 2 to 5 acres / 6 to 10 acres / More than 10 acres)
    \item How comfortable are you using mobile apps? (Linear scale 1 to 5, Very uncomfortable to Very comfortable)
\end{itemize}

\subsection{Section 3: Traditional Crop Protection}
\begin{itemize}
    \item Before using Pezego, what did you do when pests attacked your crops? (Select all that apply: Checked and picked pests by hand / Asked neighbor farmers for advice / Bought chemical sprays from a local shop without advice / Traveled to find an Extension Officer for help / Did nothing because help was too far or too expensive)
    \item Before using Pezego, how long did it usually take to get a recommendation for a pest problem? (Same day / 1 to 2 days / 3 to 7 days / More than a week / I was never able to get help)
\end{itemize}

\subsection{Section 4: App Usability \& Usefulness Grid Statements}
Farmers rate these statements on a 5-point Likert scale (1: Strongly Disagree, 5: Strongly Agree):
\begin{enumerate}
    \item[S1.] The app helps me identify pests and crop damage much faster.
    \item[S2.] The weather feature in the app helps me plan my farm work (like spraying or weeding).
    \item[S3.] The pest management advice provided by the app is useful.
    \item[S4.] Logging into the app using the SMS code was simple.
    \item[S5.] Taking pest photos and navigating the app is easy.
    \item[S6.] I plan to keep using Pezego on my farm during the next crop season.
    \item[S7.] I would recommend Pezego to other farmers in my community.
\end{enumerate}

\subsection{Section 5: AI Trust \& Officer Connection Grid Statements}
Farmers rate these statements on a 5-point Likert scale (1: Strongly Disagree, 5: Strongly Agree):
\begin{enumerate}
    \item[S1.] I trust the pest names identified by the app's camera.
    \item[S2.] I trust the farming recommendations generated by the app.
    \item[S3.] Assigning my pest report to an extension officer in the app was easy.
    \item[S4.] Knowing that an officer verified (checked) my app report makes me trust the advice more.
    \item[S5.] The button to call the assigned officer directly is helpful for urgent problems.
    \item[S6.] I received notifications when the officer checked my report, and this was helpful.
\end{enumerate}

\subsection{Section 6: Final Impact \& Feedback}
\begin{itemize}
    \item Did the app's recommendations help you save your crops from damage? (Yes, it saved most of my crops / Yes, it saved some of my crops / No, it did not save my crops)
    \item Do you feel that using Pezego helped you avoid losing money on crop damage this season? (Yes, I saved a significant amount of money / Yes, I saved a small amount of money / No, I did not save money)
    \item What was the most helpful feature of Pezego, and what was the hardest part to use? (Paragraph)
    \item Any other comments, or features you want us to add to help your farming? (Paragraph)
\end{itemize}

\section{Representative Prompt Templates for the Multi-Agent Pipeline}
\label{app:prompt_templates}
To ensure reproducibility, we present the representative prompt templates used by the core agents in the Pezego-HITL pipeline. All prompts are executed using the gpt-5.4-nano model.

\subsection{Synthesis Agent Prompt Template}
The Synthesis Agent is responsible for compiling the raw diagnostic inputs and retrieved policy documents into an actionable advisory draft. It is executed at temperature $T = 0.2$.
\begin{verbatim}
[System Prompt]
You are the Synthesis Agent in the Pezego-HITL precision agriculture framework.
Your task is to draft pest management suggestions for a farmer based on the following:
1. Pest detection output: {pest_species} (density: {pest_density})
2. Crop and growth stage: {crop_type} (stage: {growth_stage})
3. Local weather and region: {weather_conditions}, {region_district}
4. Retrieved policy documents: {policy_context}

Write a clear, structured management advice draft. You MUST adhere strictly to the 
active ingredients, dosages, and safety precautions specified in the retrieved policy 
documents. Do not recommend chemicals not explicitly listed.

[User Input]
Draft suggestions for:
- Pest: Fall Armyworm (3 larvae per plant)
- Crop: Maize (Late Vegetative, GS 35)
- Location: Ejura, Ashanti Region
- Weather: Rain forecast in 24h
\end{verbatim}

\subsection{Critique Agent Prompt Template}
The Critique Agent acts as a safety guardrail to verify that the Synthesis Agent's draft complies with regional regulations and crop safety rules. It is executed at temperature $T = 0.0$.
\begin{verbatim}
[System Prompt]
You are the Critique Agent. Your role is to audit the drafted pest management advice.
You must ensure:
1. All chemical pesticides recommended in the draft are registered in the 
   {ghana_epa_registry}.
2. The dosage recommended is within the safe thresholds for the specific crop growth stage.
3. Any regional chemical bans (e.g., restricted active ingredients in water-catchment 
   districts) are respected.

If any safety rules are violated, output a structured violation report in JSON format 
with the violation code (e.g., ERR_PST_UNAPPROVED_CROP) and rewrite instructions. If the 
draft is safe, output {"status": "APPROVED"}.

[User Input]
Audit the following draft for Maize at GS 35 in Ashanti Region:
Draft: "Apply Lambda-cyhalothrin at a rate of 150 ml/ha..."
\end{verbatim}

\section{Demographic Profiles and Traditional Practices Baseline Survey Data}
\label{app:demographic_tables}
This appendix compiles the complete demographic profiles of the ESO and farmer cohorts, along with their traditional practices and operational bottlenecks reported prior to using Pezego.

\begin{table}[htbp]
  \caption{Demographic profile of ESO respondents (Sample size $N = 30$).}
  \label{tab:demographics}
  \centering
  \small
  \begin{tabular}{llr}
    \toprule
    Demographic Variable & Category & Count (\%) \\
    \midrule
    Age & Under 25 & 4 (13.3\%) \\
        & 25--34 & 6 (20.0\%) \\
        & 35--44 & 12 (40.0\%) \\
        & 45--54 & 7 (23.3\%) \\
        & 55 or above & 1 (3.3\%) \\
    \addlinespace
    Gender & Male & 25 (83.3\%) \\
           & Female & 5 (16.7\%) \\
           & Prefer not to say & 0 (0.0\%) \\
    \addlinespace
    Highest Qualification & Certificate in Agriculture & 5 (16.7\%) \\
                          & Diploma in Agriculture & 1 (3.3\%) \\
                          & BSc in Agriculture / Extension & 19 (63.3\%) \\
                          & MSc / MPhil in Agriculture / Extension & 4 (13.3\%) \\
                          & PhD / Doctorate & 1 (3.3\%) \\
    \addlinespace
    Current ESO Role & Field Level ESO & 14 (46.7\%) \\
                     & District Level ESO & 15 (50.0\%) \\
                     & Region Level ESO & 1 (3.3\%) \\
                     & National Level ESO & 0 (0.0\%) \\
    \addlinespace
    Years of Experience & Less than 2 years & 6 (20.0\%) \\
                        & 2--5 years & 6 (20.0\%) \\
                        & 6--10 years & 8 (26.7\%) \\
                        & More than 10 years & 10 (33.3\%) \\
    \bottomrule
  \end{tabular}
\end{table}

\begin{table}[htbp]
  \caption{Traditional practices and operational bottlenecks reported by ESOs ($N = 30$; multiple selections allowed; percentages represent the proportion of respondents selecting each method).}
  \label{tab:traditional_work}
  \centering
  \small
  \begin{tabularx}{\linewidth}{lX}
    \toprule
    Traditional Diagnostic Method & Selection Frequency (\%) \\
    \midrule
    Visual check based on personal experience & 27 (90.0\%) \\
    Consulting textbooks, manuals, or reference guides & 8 (26.7\%) \\
    Sending photos to colleagues via WhatsApp/Telegram & 8 (26.7\%) \\
    Direct phone calls to senior agronomists or research institutions & 8 (26.7\%) \\
    Other & 2 (6.7\%) \\
    \midrule
    \textbf{Average Recommendation Turnaround Time} & \textbf{Percentage (\%)} \\
    \midrule
    Less than 2 hours & 13.3\% \\
    2 to 24 hours & 70.0\% \\
    2 to 3 days & 13.3\% \\
    4 to 7 days & 3.3\% \\
    More than a week & 0.0\% \\
    \bottomrule
  \end{tabularx}
\end{table}

\begin{figure}[htbp]
  \centering
  \resizebox{0.85\linewidth}{!}{
  \begin{tikzpicture}
    \begin{axis}[
        xbar,
        xlabel={Percentage of ESOs (\%)},
        xlabel style={font=\sffamily\small},
        symbolic y coords={Other, Senior Calls, WhatsApp, Books/Manuals, Visual Check},
        ytick=data,
        ytick align=outside,
        xtick align=outside,
        tick pos=left,
        axis x line*=bottom,
        axis y line*=left,
        axis line style={draw=black!60, line width=0.6pt},
        xmajorgrids=true,
        grid style={draw=black!10, line width=0.4pt, dashed},
        nodes near coords,
        nodes near coords align={horizontal},
        every node near coord/.append style={font=\sffamily\footnotesize},
        tick label style={font=\sffamily\footnotesize},
        width=10cm, height=6cm,
        xmin=0, xmax=100,
        bar width=14pt
    ]
        \addplot[fill=likertsa, draw=none] coordinates {(6.7,Other) (26.7,Senior Calls) (26.7,WhatsApp) (26.7,Books/Manuals) (90.0,Visual Check)};
    \end{axis}
  \end{tikzpicture}
  }
  \caption{Traditional pest identification methods utilised by ESOs in the field (multiple selections allowed; $N = 30$).}
  \label{fig:traditional_methods}
\end{figure}
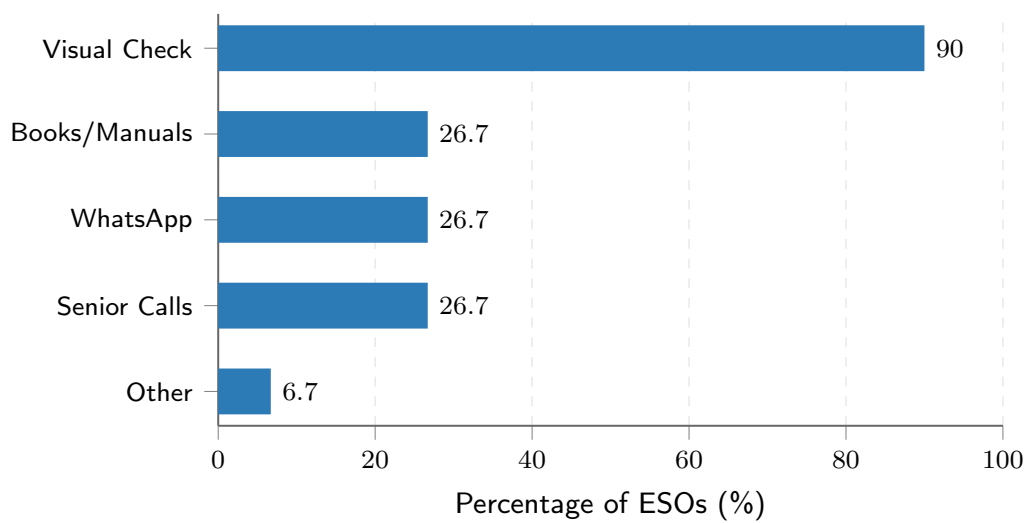

\begin{table}[htbp]
  \caption{Extension officers' estimate of crop yield loss avoided through Pezego ($N = 30$).}
  \label{tab:yield_loss}
  \centering
  \small
  \begin{tabular}{lr}
    \toprule
    Perceived Yield Loss Avoided & Selection Frequency (\%) \\
    \midrule
    Less than 5\% & 4 (13.3\%) \\
    5\% to 15\% & 8 (26.7\%) \\
    16\% to 30\% & 8 (26.7\%) \\
    31\% to 50\% & 5 (16.7\%) \\
    More than 50\% & 5 (16.7\%) \\
    \bottomrule
  \end{tabular}
\end{table}

\begin{table}[htbp]
  \caption{Demographic profile of smallholder farmer respondents (Sample size $N = 36$).}
  \label{tab:farmer_demographics}
  \centering
  \small
  \begin{tabular}{llr}
    \toprule
    Demographic Variable & Category & Count (\%) \\
    \midrule
    Age & Under 25 & 5 (13.9\%) \\
        & 25--34 & 14 (38.9\%) \\
        & 35--44 & 10 (27.8\%) \\
        & 45--54 & 3 (8.3\%) \\
        & 55 or above & 4 (11.1\%) \\
    \addlinespace
    Gender & Male & 27 (75.0\%) \\
           & Female & 9 (25.0\%) \\
    \addlinespace
    Farm Size & Less than 2 acres & 2 (5.6\%) \\
              & 2 to 5 acres & 12 (33.3\%) \\
              & 6 to 10 acres & 14 (38.9\%) \\
              & More than 10 acres & 8 (22.2\%) \\
    \bottomrule
  \end{tabular}
\end{table}

\begin{table}[htbp]
  \caption{Traditional practices and pest recommendation delays reported by farmers ($N = 36$; multiple selections allowed; percentages represent the proportion of respondents selecting each action).}
  \label{tab:farmer_traditional}
  \centering
  \small
  \begin{tabularx}{\linewidth}{lX}
    \toprule
    Traditional Crop-Protection Action & Selection Frequency (\%) \\
    \midrule
    Bought chemical sprays from a local shop without advice & 24 (66.7\%) \\
    Asked neighbor farmers for advice & 22 (61.1\%) \\
    Traveled to find an Extension Officer for help & 19 (52.8\%) \\
    Checked and picked pests by hand & 14 (38.9\%) \\
    Other & 0 (0.0\%) \\
    \midrule
    \textbf{Traditional Recommendation Turnaround Time} & \textbf{Percentage (\%)} \\
    \midrule
    Same day & 14 (38.9\%) \\
    1 to 2 days & 12 (33.3\%) \\
    3 to 7 days & 5 (13.9\%) \\
    More than a week & 4 (11.1\%) \\
    I was never able to get help & 1 (2.8\%) \\
    \bottomrule
  \end{tabularx}
\end{table}

\section{Detailed Tabular Survey Results}
\label{app:survey_tables}
This appendix contains the detailed numerical Likert scale percentage distributions and Mean/SD scores for ESOs' and farmers' evaluations of Pezego.

\subsection{ESO Likert Tables ($N = 30$)}

\begin{table}[htbp]
  \caption{Extension officer agreement ratings on Pezego usability and usefulness ($N = 30$; reliability $\alpha = 0.967$).}
  \label{tab:usability}
  \centering
  \small
  \begin{tabularx}{\linewidth}{lXXXXXX}
    \toprule
    Statement & 1 (\%) & 2 (\%) & 3 (\%) & 4 (\%) & 5 (\%) & Mean (SD) \\
    \midrule
    S1 & 10.0\% & 0.0\% & 10.0\% & 30.0\% & 50.0\% & 4.10 (1.24) \\
    S2 & 6.7\% & 6.7\% & 3.3\% & 36.7\% & 46.7\% & 4.10 (1.18) \\
    S3 & 6.7\% & 0.0\% & 10.0\% & 46.7\% & 36.7\% & 4.07 (1.05) \\
    S4 & 3.3\% & 3.3\% & 13.3\% & 46.7\% & 33.3\% & 4.03 (0.96) \\
    S5 & 10.0\% & 6.7\% & 6.7\% & 36.7\% & 40.0\% & 3.90 (1.30) \\
    S6 & 6.7\% & 3.3\% & 13.3\% & 26.7\% & 50.0\% & 4.10 (1.18) \\
    S7 & 6.7\% & 3.3\% & 10.0\% & 16.7\% & 63.3\% & 4.27 (1.20) \\
    \bottomrule
  \end{tabularx}
\end{table}

\begin{table}[htbp]
  \caption{ESOs' evaluation of AI diagnostics, explanations, and verification safety ($N = 30$; reliability $\alpha = 0.885$).}
  \label{tab:ai_quality}
  \centering
  \small
  \begin{tabularx}{\linewidth}{lXXXXXX}
    \toprule
    Statement & 1 (\%) & 2 (\%) & 3 (\%) & 4 (\%) & 5 (\%) & Mean (SD) \\
    \midrule
    S1 & 3.3\% & 0.0\% & 13.3\% & 33.3\% & 50.0\% & 4.27 (0.94) \\
    S2 & 3.3\% & 3.3\% & 10.0\% & 36.7\% & 46.7\% & 4.20 (1.00) \\
    S3 & 3.3\% & 6.7\% & 10.0\% & 46.7\% & 33.3\% & 4.00 (1.02) \\
    S4 & 3.3\% & 3.3\% & 10.0\% & 23.3\% & 60.0\% & 4.33 (1.03) \\
    S5 & 3.3\% & 3.3\% & 10.0\% & 36.7\% & 46.7\% & 4.20 (1.00) \\
    S6 & 6.7\% & 23.3\% & 23.3\% & 20.0\% & 26.7\% & 3.37 (1.30) \\
    \bottomrule
  \end{tabularx}
\end{table}

\begin{table}[htbp]
  \caption{ESOs' ratings of communication, assignment, and community features ($N = 30$; reliability $\alpha = 0.959$).}
  \label{tab:communication}
  \centering
  \small
  \begin{tabularx}{\linewidth}{lXXXXXX}
    \toprule
    Statement & 1 (\%) & 2 (\%) & 3 (\%) & 4 (\%) & 5 (\%) & Mean (SD) \\
    \midrule
    S1 & 3.3\% & 13.3\% & 10.0\% & 30.0\% & 43.3\% & 3.97 (1.19) \\
    S2 & 3.3\% & 6.7\% & 10.0\% & 30.0\% & 50.0\% & 4.17 (1.09) \\
    S3 & 3.3\% & 10.0\% & 6.7\% & 23.3\% & 56.7\% & 4.20 (1.16) \\
    S4 & 3.3\% & 6.7\% & 13.3\% & 26.7\% & 50.0\% & 4.13 (1.11) \\
    S5 & 6.7\% & 6.7\% & 16.7\% & 43.3\% & 26.7\% & 3.77 (1.14) \\
    S6 & 6.7\% & 3.3\% & 13.3\% & 30.0\% & 46.7\% & 4.07 (1.17) \\
    \bottomrule
  \end{tabularx}
\end{table}

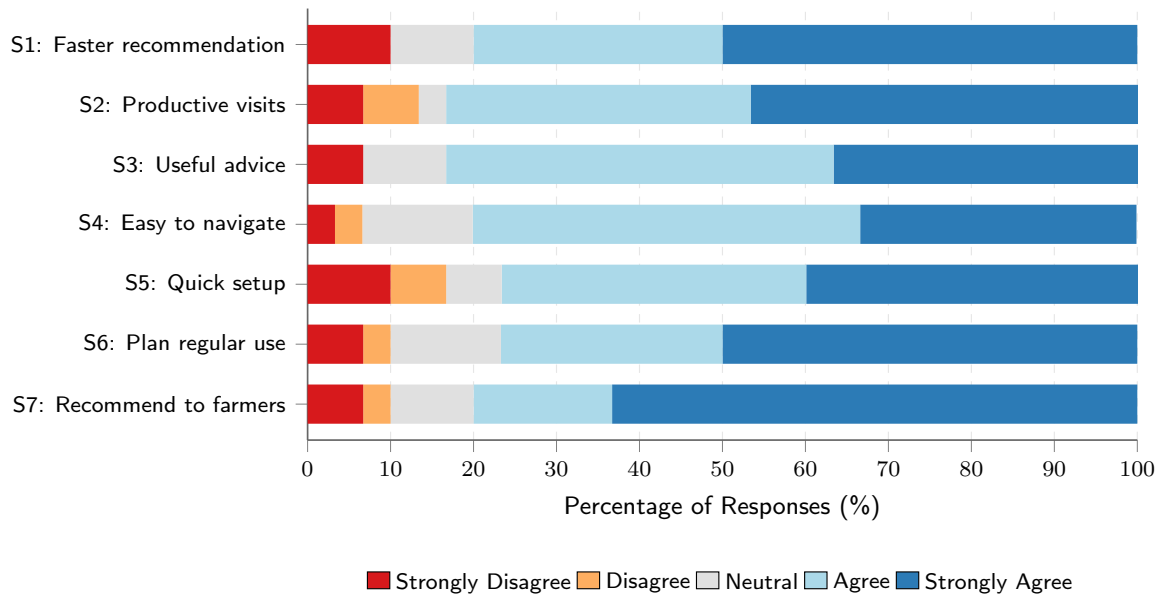
\begin{figure}[htbp]
  \centering
  \resizebox{0.95\linewidth}{!}{
  \begin{tikzpicture}
    \begin{axis}[
        xbar stacked,
        legend style={at={(0.5,-0.28)}, anchor=north, legend columns=-1, font=\sffamily\footnotesize, draw=none, fill=none},
        xlabel={Percentage of Responses (\%)},
        xlabel style={font=\sffamily\small},
        ytick={1,2,3,4,5,6,7},
        yticklabels={
            {S7: Recommend to farmers},
            {S6: Plan regular use},
            {S5: Quick setup},
            {S4: Easy to navigate},
            {S3: Useful advice},
            {S2: Productive visits},
            {S1: Faster recommendation}
        },
        ytick align=outside,
        xtick align=outside,
        tick pos=left,
        axis x line*=bottom,
        axis y line*=left,
        axis line style={draw=black!60, line width=0.6pt},
        xmajorgrids=true,
        grid style={draw=black!10, line width=0.4pt, dashed},
        tick label style={font=\sffamily\footnotesize},
        xmin=0, xmax=100,
        width=12cm, height=7cm,
        bar width=14pt
    ]
    % Strongly Disagree (1)
    \addplot[fill=likertsd, draw=none] coordinates {(10.0,7) (6.7,6) (6.7,5) (3.3,4) (10.0,3) (6.7,2) (6.7,1)};
    % Disagree (2)
    \addplot[fill=likertd, draw=none] coordinates {(0.0,7) (6.7,6) (0.0,5) (3.3,4) (6.7,3) (3.3,2) (3.3,1)};
    % Neutral (3)
    \addplot[fill=likertn, draw=none] coordinates {(10.0,7) (3.3,6) (10.0,5) (13.3,4) (6.7,3) (13.3,2) (10.0,1)};
    % Agree (4)
    \addplot[fill=likerta, draw=none] coordinates {(30.0,7) (36.7,6) (46.7,5) (46.7,4) (36.7,3) (26.7,2) (16.7,1)};
    % Strongly Agree (5)
    \addplot[fill=likertsa, draw=none] coordinates {(50.0,7) (46.7,6) (36.7,5) (33.3,4) (40.0,3) (50.0,2) (63.3,1)};
    \legend{Strongly Disagree, Disagree, Neutral, Agree, Strongly Agree}
    \end{axis}
  \end{tikzpicture}
  }
  \caption{Distribution of Likert scale ratings for App Usability and Usefulness (S1--S7 correspond to Table \ref{tab:usability}; $N = 30$).}
  \label{fig:usability_likert}
\end{figure}

\begin{figure}[htbp]
  \centering
  \resizebox{0.95\linewidth}{!}{
  \begin{tikzpicture}
    \begin{axis}[
        xbar stacked,
        legend style={at={(0.5,-0.28)}, anchor=north, legend columns=-1, font=\sffamily\footnotesize, draw=none, fill=none},
        xlabel={Percentage of Responses (\%)},
        xlabel style={font=\sffamily\small},
        ytick={1,2,3,4,5,6},
        yticklabels={
            {S6: Noticed AI mistakes},
            {S5: Verification confidence},
            {S4: Important to modify},
            {S3: Safe/practical advice},
            {S2: Clear explanations},
            {S1: Correct AI recommendation}
        },
        ytick align=outside,
        xtick align=outside,
        tick pos=left,
        axis x line*=bottom,
        axis y line*=left,
        axis line style={draw=black!60, line width=0.6pt},
        xmajorgrids=true,
        grid style={draw=black!10, line width=0.4pt, dashed},
        tick label style={font=\sffamily\footnotesize},
        xmin=0, xmax=100,
        width=12cm, height=7.5cm,
        bar width=14pt
    ]
    % Strongly Disagree (1)
    \addplot[fill=likertsd, draw=none] coordinates {(3.3,6) (3.3,5) (3.3,4) (3.3,3) (3.3,2) (6.7,1)};
    % Disagree (2)
    \addplot[fill=likertd, draw=none] coordinates {(0.0,6) (3.3,5) (6.7,4) (3.3,3) (3.3,2) (23.3,1)};
    % Neutral (3)
    \addplot[fill=likertn, draw=none] coordinates {(13.3,6) (10.0,5) (10.0,4) (10.0,3) (10.0,2) (23.3,1)};
    % Agree (4)
    \addplot[fill=likerta, draw=none] coordinates {(33.3,6) (36.7,5) (46.7,4) (23.3,3) (36.7,2) (20.0,1)};
    % Strongly Agree (5)
    \addplot[fill=likertsa, draw=none] coordinates {(50.0,6) (46.7,5) (33.3,4) (60.0,3) (46.7,2) (26.7,1)};
    \legend{Strongly Disagree, Disagree, Neutral, Agree, Strongly Agree}
    \end{axis}
  \end{tikzpicture}
  }
  \caption{Distribution of Likert scale ratings for AI Diagnostics and Verification Safety (S1--S6 correspond to Table \ref{tab:ai_quality}; $N = 30$).}
  \label{fig:ai_diagnostics_likert}
\end{figure}
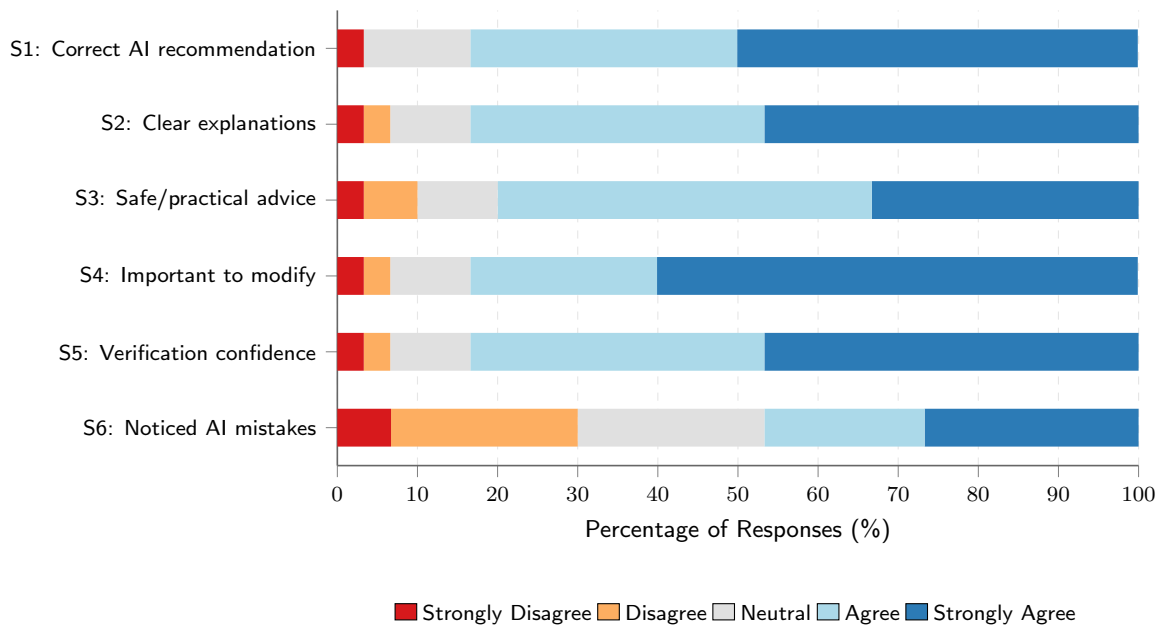

\begin{figure}[htbp]
  \centering
  \resizebox{0.95\linewidth}{!}{
  \begin{tikzpicture}
    \begin{axis}[
        xbar stacked,
        legend style={at={(0.5,-0.85)}, anchor=north, legend columns=-1, font=\sffamily\footnotesize, draw=none, fill=none},
        xlabel={Percentage of ESOs (\%)},
        xlabel style={font=\sffamily\small},
        ytick=\empty,
        yticklabels={},
        axis x line*=bottom,
        axis y line=none,
        tick pos=left,
        xtick align=outside,
        tick label style={font=\sffamily\footnotesize},
        xmin=0, xmax=100,
        ymin=0.5, ymax=1.5,
        width=12cm, height=3.2cm,
        bar width=22pt,
        enlarge y limits={abs=0.5cm}
    ]
    % <5%
    \addplot[fill=likertsd, draw=none] coordinates {(13.3,1)};
    % 5-15%
    \addplot[fill=likertd, draw=none] coordinates {(26.7,1)};
    % 16-30%
    \addplot[fill=likertn, draw=none] coordinates {(26.7,1)};
    % 31-50%
    \addplot[fill=likerta, draw=none] coordinates {(16.7,1)};
    % >50%
    \addplot[fill=likertsa, draw=none] coordinates {(16.7,1)};
    \legend{<5\%, 5--15\%, 16--30\%, 31--50\%, >50\%}
    \end{axis}
  \end{tikzpicture}
  }
  \caption{ESOs' estimates of the crop yield loss avoided by farmers using Pezego ($N = 30$).}
  \label{fig:yield_loss_avoided}
\end{figure}
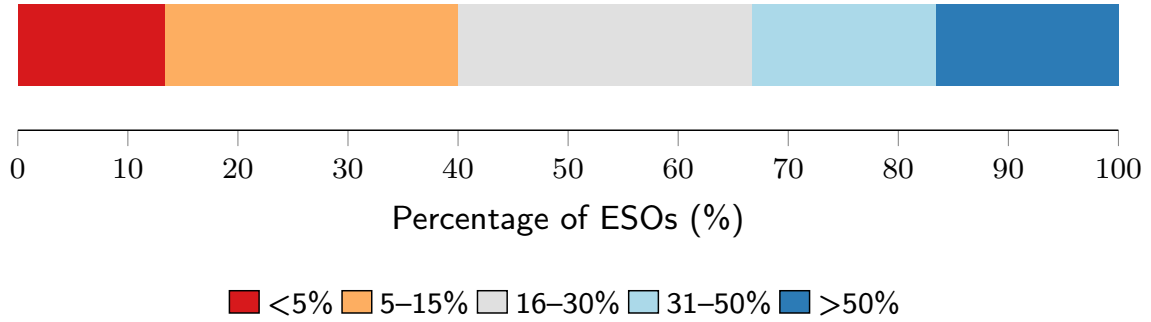

\subsection{Farmer Survey Likert Tables ($N = 36$)}

\begin{table}[htbp]
  \caption{Farmer agreement ratings on Pezego usability and usefulness ($N = 36$; reliability $\alpha = 0.949$).}
  \label{tab:farmer_usability}
  \centering
  \small
  \begin{tabularx}{\linewidth}{lXXXXXX}
    \toprule
    Statement & 1 (\%) & 2 (\%) & 3 (\%) & 4 (\%) & 5 (\%) & Mean (SD) \\
    \midrule
    S1 & 0.0\% & 5.6\% & 11.1\% & 19.4\% & 63.9\% & 4.42 (0.91) \\
    S2 & 0.0\% & 0.0\% & 13.9\% & 27.8\% & 58.3\% & 4.44 (0.73) \\
    S3 & 2.8\% & 5.6\% & 5.6\% & 25.0\% & 61.1\% & 4.36 (1.02) \\
    S4 & 8.3\% & 2.8\% & 11.1\% & 22.2\% & 55.6\% & 4.14 (1.25) \\
    S5 & 0.0\% & 0.0\% & 13.9\% & 25.0\% & 61.1\% & 4.47 (0.74) \\
    S6 & 0.0\% & 5.6\% & 8.3\% & 25.0\% & 61.1\% & 4.42 (0.87) \\
    S7 & 0.0\% & 0.0\% & 13.9\% & 19.4\% & 66.7\% & 4.53 (0.74) \\
    \bottomrule
  \end{tabularx}
\end{table}

\begin{table}[htbp]
  \caption{Farmers' trust and connection ratings on Pezego diagnostics ($N = 36$; reliability $\alpha = 0.944$).}
  \label{tab:farmer_trust}
  \centering
  \small
  \begin{tabularx}{\linewidth}{lXXXXXX}
    \toprule
    Statement & 1 (\%) & 2 (\%) & 3 (\%) & 4 (\%) & 5 (\%) & Mean (SD) \\
    \midrule
    S1 & 2.8\% & 0.0\% & 2.8\% & 52.8\% & 41.7\% & 4.31 (0.79) \\
    S2 & 0.0\% & 2.8\% & 5.6\% & 50.0\% & 41.7\% & 4.31 (0.71) \\
    S3 & 0.0\% & 5.6\% & 2.8\% & 38.9\% & 52.8\% & 4.39 (0.80) \\
    S4 & 0.0\% & 0.0\% & 5.6\% & 38.9\% & 55.6\% & 4.50 (0.61) \\
    S5 & 0.0\% & 0.0\% & 5.6\% & 33.3\% & 61.1\% & 4.56 (0.61) \\
    S6 & 0.0\% & 2.8\% & 2.8\% & 33.3\% & 61.1\% & 4.53 (0.70) \\
    \bottomrule
  \end{tabularx}
\end{table}

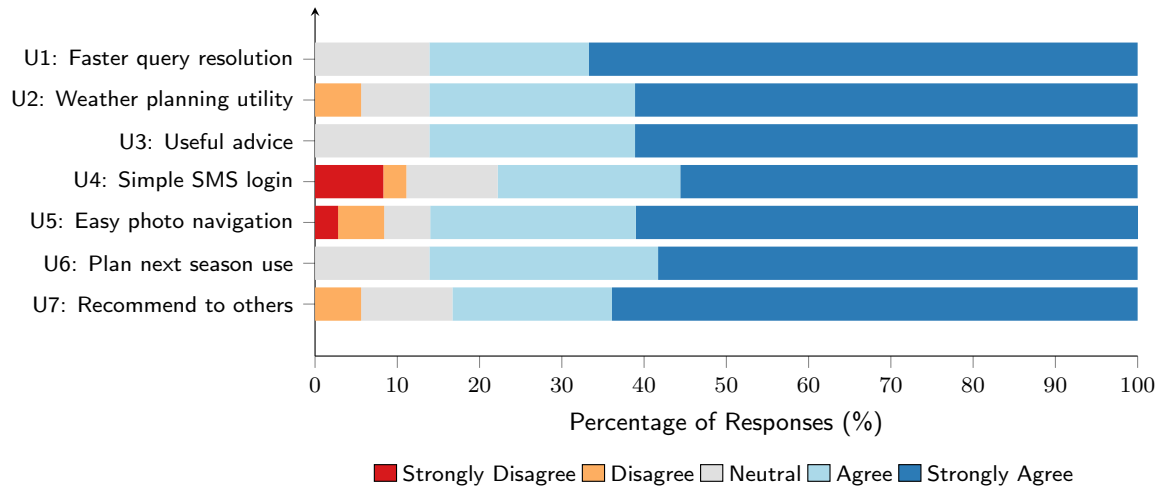
\begin{figure}[htbp]
  \centering
  \resizebox{0.95\linewidth}{!}{
  \begin{tikzpicture}
    \begin{axis}[
        xbar stacked,
        legend style={at={(0.5,-0.28)}, anchor=north, legend columns=-1, font=\sffamily\footnotesize, draw=none, fill=none},
        xlabel={Percentage of Responses (\%)},
        xlabel style={font=\sffamily\small},
        ytick={1,2,3,4,5,6,7},
        yticklabels={
            {U7: Recommend to others},
            {U6: Plan next season use},
            {U5: Easy photo navigation},
            {U4: Simple SMS login},
            {U3: Useful advice},
            {U2: Weather planning utility},
            {U1: Faster query resolution}
        },
        axis x line*=bottom,
        axis y line=left,
        tick pos=left,
        xtick align=outside,
        ytick align=outside,
        tick label style={font=\sffamily\footnotesize},
        xmin=0, xmax=100,
        ymin=0.5, ymax=7.5,
        width=12cm, height=6.0cm,
        bar width=12pt,
        enlarge y limits={abs=0.4cm}
    ]
    % Strongly Disagree (1)
    \addplot[fill=likertsd, draw=none] coordinates {(0.0,7) (0.0,6) (0.0,5) (8.3,4) (2.8,3) (0.0,2) (0.0,1)};
    % Disagree (2)
    \addplot[fill=likertd, draw=none] coordinates {(0.0,7) (5.6,6) (0.0,5) (2.8,4) (5.6,3) (0.0,2) (5.6,1)};
    % Neutral (3)
    \addplot[fill=likertn, draw=none] coordinates {(13.9,7) (8.3,6) (13.9,5) (11.1,4) (5.6,3) (13.9,2) (11.1,1)};
    % Agree (4)
    \addplot[fill=likerta, draw=none] coordinates {(19.4,7) (25.0,6) (25.0,5) (22.2,4) (25.0,3) (27.8,2) (19.4,1)};
    % Strongly Agree (5)
    \addplot[fill=likertsa, draw=none] coordinates {(66.7,7) (61.1,6) (61.1,5) (55.6,4) (61.1,3) (58.3,2) (63.9,1)};
    \legend{Strongly Disagree, Disagree, Neutral, Agree, Strongly Agree}
    \end{axis}
  \end{tikzpicture}
  }
  \caption{Smallholder farmers' agreement ratings on Pezego usability and usefulness (U1--U7; $N = 36$; reliability $\alpha = 0.949$).}
  \label{fig:farmer_usability_likert}
\end{figure}

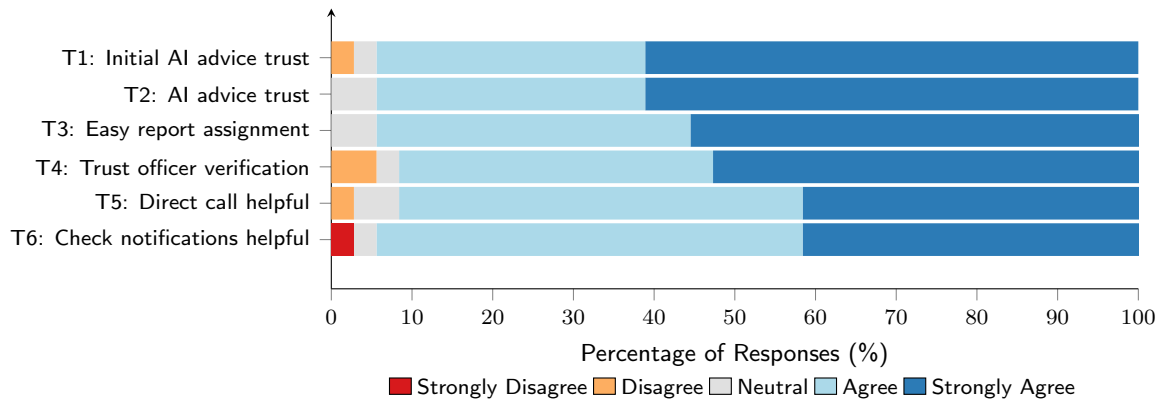
\begin{figure}[htbp]
  \centering
  \resizebox{0.95\linewidth}{!}{
  \begin{tikzpicture}
    \begin{axis}[
        xbar stacked,
        legend style={at={(0.5,-0.28)}, anchor=north, legend columns=-1, font=\sffamily\footnotesize, draw=none, fill=none},
        xlabel={Percentage of Responses (\%)},
        xlabel style={font=\sffamily\small},
        ytick={1,2,3,4,5,6},
        yticklabels={
            {T6: Check notifications helpful},
            {T5: Direct call helpful},
            {T4: Trust officer verification},
            {T3: Easy report assignment},
            {T2: AI advice trust},
            {T1: Initial AI advice trust}
        },
        axis x line*=bottom,
        axis y line=left,
        tick pos=left,
        xtick align=outside,
        ytick align=outside,
        tick label style={font=\sffamily\footnotesize},
        xmin=0, xmax=100,
        ymin=0.5, ymax=6.5,
        width=12cm, height=5.2cm,
        bar width=12pt,
        enlarge y limits={abs=0.4cm}
    ]
    % Strongly Disagree (1)
    \addplot[fill=likertsd, draw=none] coordinates {(0.0,6) (0.0,5) (0.0,4) (0.0,3) (0.0,2) (2.8,1)};
    % Disagree (2)
    \addplot[fill=likertd, draw=none] coordinates {(2.8,6) (0.0,5) (0.0,4) (5.6,3) (2.8,2) (0.0,1)};
    % Neutral (3)
    \addplot[fill=likertn, draw=none] coordinates {(2.8,6) (5.6,5) (5.6,4) (2.8,3) (5.6,2) (2.8,1)};
    % Agree (4)
    \addplot[fill=likerta, draw=none] coordinates {(33.3,6) (33.3,5) (38.9,4) (38.9,3) (50.0,2) (52.8,1)};
    % Strongly Agree (5)
    \addplot[fill=likertsa, draw=none] coordinates {(61.1,6) (61.1,5) (55.6,4) (52.8,3) (41.7,2) (41.7,1)};
    \legend{Strongly Disagree, Disagree, Neutral, Agree, Strongly Agree}
    \end{axis}
  \end{tikzpicture}
  }
  \caption{Smallholder farmers' agreement ratings on AI trust and officer connection (T1--T6; $N = 36$; reliability $\alpha = 0.944$).}
  \label{fig:farmer_trust_likert}
\end{figure}

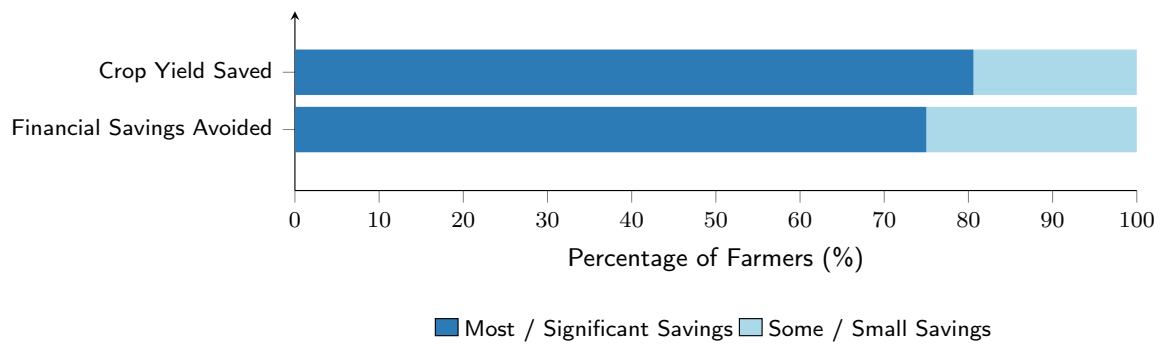
\begin{figure}[htbp]
  \centering
  \resizebox{0.95\linewidth}{!}{
  \begin{tikzpicture}
    \begin{axis}[
        xbar stacked,
        legend style={at={(0.5,-0.65)}, anchor=north, legend columns=-1, font=\sffamily\footnotesize, draw=none, fill=none},
        xlabel={Percentage of Farmers (\%)},
        xlabel style={font=\sffamily\small},
        ytick={1,2},
        yticklabels={
            {Financial Savings Avoided},
            {Crop Yield Saved}
        },
        axis x line*=bottom,
        axis y line=left,
        tick pos=left,
        xtick align=outside,
        ytick align=outside,
        tick label style={font=\sffamily\footnotesize},
        xmin=0, xmax=100,
        ymin=0.5, ymax=2.5,
        width=12cm, height=3.8cm,
        bar width=16pt,
        enlarge y limits={abs=0.4cm}
    ]
    % Most/Significant (likertsa)
    \addplot[fill=likertsa, draw=none] coordinates {(75.0,1) (80.6,2)};
    % Some/Small (likerta)
    \addplot[fill=likerta, draw=none] coordinates {(25.0,1) (19.4,2)};
    \legend{Most / Significant Savings, Some / Small Savings}
    \end{axis}
  \end{tikzpicture}
  }
  \caption{Smallholder farmers' estimates of crop yield protection and financial savings avoided by using Pezego ($N = 36$).}
  \label{fig:farmer_savings}
\end{figure}

\end{document}